UNIVERSITY OF CALIFORNIA, MERCED

**PrivateJobMatch: A Privacy-Oriented Deferred Multi-Match Recommender System for Stable Employment**

in

Electrical Engineering and Computer Science

by

Amar Saini

Committee in charge:

    Professor Florin Rusu, Chair
    Assistant Professor Andrew Johnston
    Professor Marcelo Kallmann
    Professor Alberto Cerpa

2019

The Thesis of Amar Saini is approved, and it is acceptable
in quality and for publication on microfilm and electronically:

\_\_\_\_\_\_\_\_\_\_\_\_\_\_\_\_\_\_\_\_\_\_\_\_\_\_\_\_\_\_\_\_\_\_\_\_\_\_\_\_\_\_\_\_\_\_\_\_\_\_\_\_\_\_\_\_\_\_\_\_\_\_

\_\_\_\_\_\_\_\_\_\_\_\_\_\_\_\_\_\_\_\_\_\_\_\_\_\_\_\_\_\_\_\_\_\_\_\_\_\_\_\_\_\_\_\_\_\_\_\_\_\_\_\_\_\_\_\_\_\_\_\_\_\_

\_\_\_\_\_\_\_\_\_\_\_\_\_\_\_\_\_\_\_\_\_\_\_\_\_\_\_\_\_\_\_\_\_\_\_\_\_\_\_\_\_\_\_\_\_\_\_\_\_\_\_\_\_\_\_\_\_\_\_\_\_\_

\_\_\_\_\_\_\_\_\_\_\_\_\_\_\_\_\_\_\_\_\_\_\_\_\_\_\_\_\_\_\_\_\_\_\_\_\_\_\_\_\_\_\_\_\_\_\_\_\_\_\_\_\_\_\_\_\_\_\_\_\_\_

Chair

University of California, Merced

2019

# Table of Contents

## Declaration

I hereby declare that no portion of the work referred to in this Project Thesis has been submitted in support of an application for another degree or qualification in this or any other university or institute of learning. If any act of plagiarism is found, I am fully responsible for every disciplinary action taken against me, depending upon the seriousness of the proven offence.

## Acknowledgements

This is for my jaaniman, Twinkle Mistry. Without her, none of this wouldn't have been possible. She is what encouraged me into graduate studies and will forever be my role model. Her love and support are what gave me the dedication for this project thesis. I'd also like to dedicate this thesis to both of my parents, and my advisor Florin Rusu, who provided strong guidance and taught me various important concepts in Computer Science.

# Abstract

**PrivateJobMatch: A Privacy-Oriented Deferred Multi-Match Recommender System for Stable Employment**

by

Amar Saini

Master of Science

in

Electrical Engineering and Computer Science

University of California, Merced

Coordination failure reduces match quality among employers and candidates in the job market, resulting in a large number of unfilled positions and/or unstable, short-term employment. Centralized job search engines provide a platform that connects directly employers with job-seekers. However, they require users to disclose a significant amount of personal data, i.e., build a user profile, in order to provide meaningful recommendations. In this paper, we present PrivateJobMatch -- a privacy-oriented deferred multi-match recommender system -- which generates stable pairings while requiring users to provide only a partial ranking of their preferences. PrivateJobMatch explores a series of adaptations of the game-theoretic Gale-Shapley deferred-acceptance algorithm which combine the flexibility of decentralized markets with the intelligence of centralized matching. We identify the shortcomings of the original algorithm when applied to a job market and propose novel solutions that rely on machine learning techniques. Experimental results on real and synthetic data confirm the benefits of the proposed algorithms across several quality measures. Over the past year, we have implemented a PrivateJobMatch prototype and deployed it in an active job market economy. Using the gathered real-user preference data, we find that the match-recommendations are superior to a typical decentralized job market---while requiring only a partial ranking of the user preferences.

# Chapter 1

# Introduction

## 1.1 The Modern-Day Recommender System for Employment

In today's decentralized job market, recommender systems have been implemented to favor candidates individually; by trying to recommend jobs to candidates on a best-fit basis. "Best-fit basis" refers to finding job recommendations that match a candidate's interests as best as possible. There are explicit and implicit methods to evaluate and capture a candidate's interests for job recommendations. Most modern-day recommender systems follow a system architecture shown in Figure 1.1.

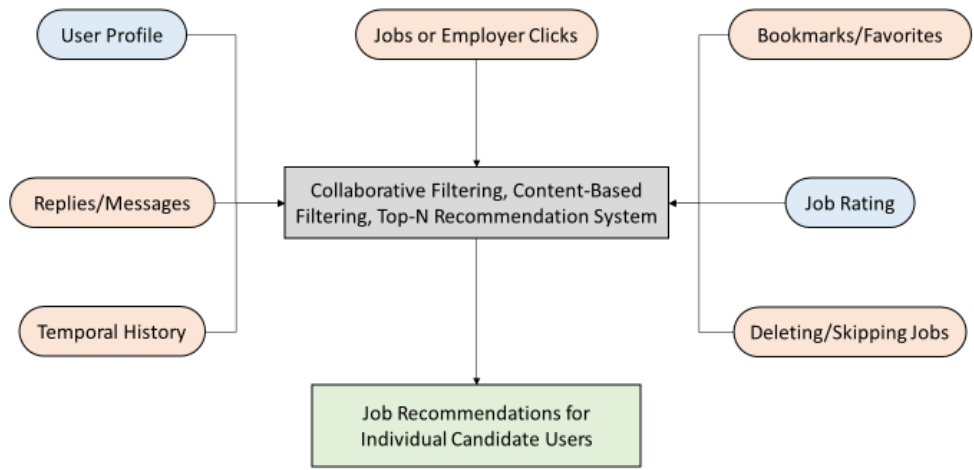

Figure 1.1: Modern Job Recommender System Architectures

In Figure 1.1, the light blue boxes represent explicit user data. This is data that the user is knowingly providing, such as building a user profile or rating a job. Light orange boxes refer to implicit user data. This type of data is gathered from the user without them being aware. Examples of this include clicking on jobs, responding to interested employers, bookmarking a job for later, temporal-based user activity, or even scrolling past uninteresting jobs. Popular modern recommender systems, such as *LinkedIn* or *Indeed*, collect their data using these explicit and implicit methods. Once the data has been acquired, collaborative filtering or content-based filtering is applied to achieve a Top-N jobs recommendation for the individual candidate user. In short, collaborative filtering is a technique that's used to find recommendations based on data that share similar features. While there is much research on what is the best explicit or implicit features to acquire, or what is the best machine learning technique to use for collaborative filtering; this type of recommender system has many drawbacks.

The first drawback of these modern recommender systems is that the user has to provide lots of explicit data in order to receive accurate recommendations. Using *LinkedIn* as an example, a candidate has to create a user profile that requires lots of information, such as a resume, education, job experience, past projects, skills, courses, accomplishments, interests, and even a personal background. This puts more work on the candidate in order for this recommender system to work. On top of the explicitly provided data, *LinkedIn* will have to implement all the methods needed to gather the implicit data, such as tracking user clicks, history, messages with employers, bookmarks, and even correctly parsing user resumes and profile backgrounds for important keywords. After successfully gathering all the data, feature selection for the collaborative filtering will be another important task that must be carefully designed. These recommender systems also eliminate the privacy of the users by requiring them to provide information for a profile.

The main drawback of this modern-day recommender system is that it encourages a decentralized job market. When hiring, employers can err either by pursuing candidates that will have better offers, which reduces options and risks a failed search, or neglect preferred, attainable candidates in favor of those that seem more likely to accept an offer. Put simply, employers justifiably struggle to identify the best, attainable candidate. That failure is costly for employers as well as for job-seekers. Addressing coordination failure may significantly help both job seekers and employers identify well-suited matches, betting the job market outcome. Using *LinkedIn* as an example again, there can be many candidates that have the same recommended job offer, but the employer may only need one. These recommender systems are encouraging a decentralized job market because it offers many job recommendations to candidates, but many of them are the same recommendations across different candidates. On the other hand, the employers may only need one of these candidates who received the same job recommendation, leaving the rest of the candidates still competition with each other for employment.

## 1.2 A Recommender System Driven by the DAA

Centralized matching markets, like those for medical residency, resolve this coordination failure by eliciting preference rankings from both sides of the market and playing out those preferences to find stable matches between candidates and employers using what is now known as the Gale-Shapley deferred-acceptance algorithm (DAA) [17, 18]. The resulting stable matches perform better than decentralized matches [19], and they improve the welfare of both sides of the market, being weakly Pareto efficient [15]. Unfortunately, these benefits are not widely realized in the labor market because they require an ex ante commitment to the outcome of the centralized match. We propose a solution that marries the flexibility of decentralized markets and the wise pairing of centralized matching by adapting the DAA into a recommender system that generates a ranked list of candidates for employers, in the order of expected match quality.

The Deferred Acceptance Algorithm (DAA) is used to solve the problem of finding a stable matching between two ordered preference datasets. This problem that the DAA solves is also known as the stable marriage problem, which states that given N men and N women, where each person has ranked all members of the opposite sex in order of preference, marry the men and women together such that there are no two people of opposite sex who would both rather have each other than their current partners. If there are no such people, all the marriages are "stable".

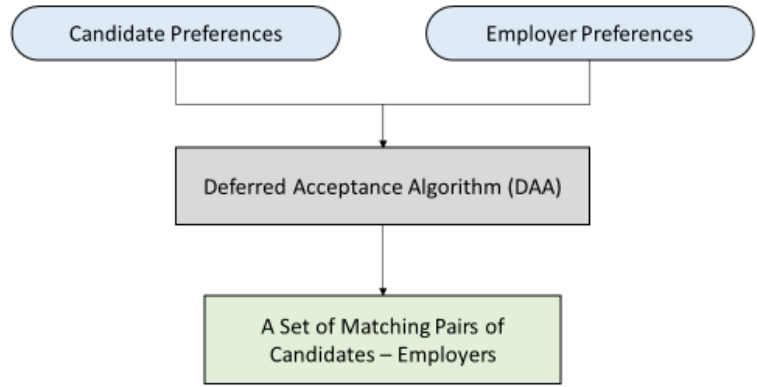

Figure 1.2: DAA Architecture

    Essentially, the DAA offers a solution to find a one-to-one matching between two sets of data. Some common examples of its application would be to match: men and women on a dating website, job seekers and employers (job market), or even schools and students. The DAA is unique because it offers a set of one-to-one matching pairs that is *stable*. For example, if we are trying to match job seekers (candidates) to jobs (employers), we can use the DAA to find a *stable* set of matches. As shown in Figure 1.2, the DAA requires two things as input, 1) the candidates' preferences of each employer, and 2) the employers' preferences of each candidate. The DAA will use these preferences as a measure of how "good" a match between a candidate and employer really is, and to see if a better match exists.

    Rather than a recommender system that is based on excessive amounts of explicit and implicit user data, we propose a recommender system that is driven by light-weight and private explicit data: bipartisan rankings. We implemented a recommender system that revolves around the DAA, and outputs stable pairs of matches (recommendations), rather than job recommendations that are best-fit for individual candidates. Figure 1.3 shows a recommender system architecture that is driven by the DAA.

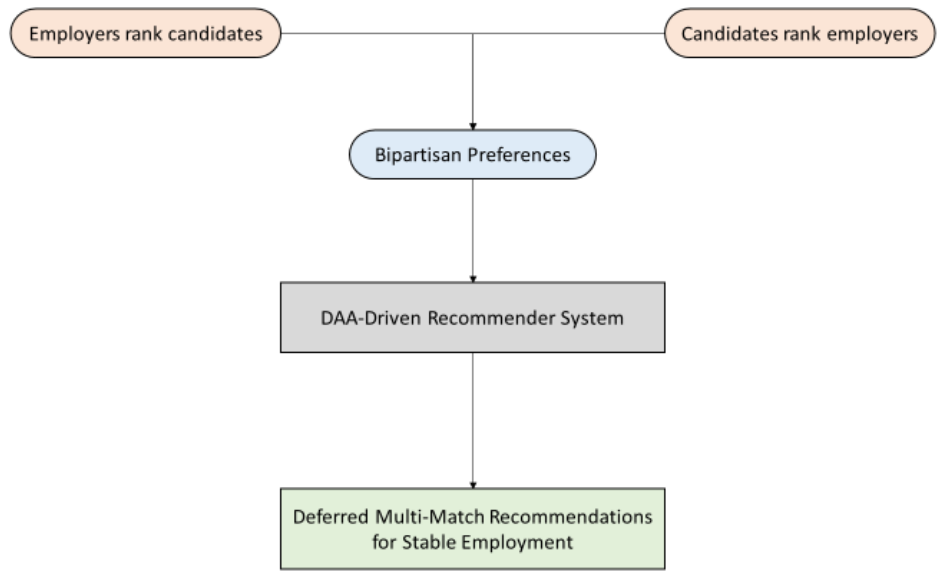

Figure 1.3: A DAA-Driven Recommender System Architecture

The novelty of this approach to job recommender systems is that it utilizes the benefits of the DAA, which is stability. A DAA-driven recommender system optimizes the stability of the job market by taking a centralized approach to ensure that candidates and employers are stably matched with each other. While modern-day recommender systems maximize a candidate's recommended job list, it doesn't provide any stability to the job market. If anything, modern-day recommender systems encourage a decentralized job market. Although stable one-to-one matches benefits the job market as a whole by reducing vacancy, it also benefits specific employers by helping reduce coordination err, which in-turn helps reduce the amount of resources (time, effort, and money) spent on the hiring process.

As with any recommender system, there are some consequences, specifically with our current adaption of the DAA into this new type of recommender system. We sought to improve this DAA-driven recommender system with the use of novel machine learning approaches, such as low-rank matrix factorization, or non-negative matrix factorization [20]. Using both synthetic and real preference data, we can test how the job market will improve based on our recommender system with the help of a machine learning approach. Through a job market simulation, we found that the match quality and job vacancy is better with our recommender system when compared to the results of a decentralized market. Our recommender system helps reduce job vacancy, and we found that it can be further improved with machine learning.

## 1.3 Contributions

A recommender system built on top the DAA, rather than one based on collaborative filtering, has many benefits. Our recommender system alleviates the work of the user, if they prefer to not provide private explicit data. There is no need to provide an extensive amount of information to acquire job recommendations. This new recommender system only requires bipartisan preferences as input, which means there is no need for implementing methods to obtain implicit data. Our recommender system also eliminates the need for feature selection, as we don't have so many different types of data, or ways to combine them. There isn't a need to decide what is the best flavor of collaborative filtering or content-based filtering, as this centralized recommender system is driven by the DAA. Since the user doesn't need to provide any personal information besides the bipartisan preferences, we ensure that the recommender system is privacy-oriented. With the help machine learning, we can improve our recommender system in terms of match accuracy and vacancy.

These advantages make a recommender system simpler to implement and manage, and it's more appealing to users as they won't have to sacrifice too much time creating a user profile. This type of recommender system also helps stabilize the job market, as we are finding stable one-to-one matches for recommendations. In modern-day job recommender systems, a job may appear as a recommendation for multiple candidates, which may result in many of the candidates not finding employment due to them competing over the same job. Our recommender system helps better the job market by finding multiple stable pairs of candidates and employers that are best for each other, based off of their bipartisan preferences. A stable job market helps employers reduce err since they won't be pursuing candidates that have better offers. Stable matches will help employers identify the best, attainable candidate, which reduces the risk of a failed search. A recommender system driven by the DAA addresses coordination failure which significantly helps both job seekers and employers identify well-suited matches. Modern recommender systems don't maximize job market stability because they are designed to help each candidate find jobs individually, while our proposed centralized recommender system takes into account the big picture and finds optimal pairs of candidates and employers to better the job market. Below are are the contributions made:

- We incorporate the DAA into a recommender system that finds multiple stable matches between candidates and employers as recommendations for employment.
- We designed several adaptations of the DAA for job recommendation, including MMDAA, LMF-MMDAA, and Mixed MMDAA.
- We implemented machine learning techniques on the bipartisan preferences, such as Low-Rank Matrix Factorization (LMF) or Non-Negative Matrix Factorization (NNMF), to enhance the resulting matches from the MMDAA, which we denote as LMF-MMDAA.
- Our proposed Mixed MMDAA is the most efficient algorithm variation of the DAA in terms of match accuracy and vacancy, which uses results from both MMDAA and LMF-MMDAA.
- We run experiments over real and synthetic data with each adapted algorithm and we show that the Mixed MMDAA is the optimal algorithm based on our newly defined measures: *Displacement*, *Withholdings*, and *Vacancy*.
- Our novel recommender system was deployed in a real application, EconMatch, with real users and we proved that it performs much better than a standard decentralized job market simulation, achieving our goal of maximizing job market stability.

# Chapter 2

# Related Work

## 2.1    Past Job Recommender Systems

In previous paper submissions for the RecSys 2016 Challenge, there was much research regarding job recommender systems using collaborative filtering on explicit and implicit user data. Some research focused on using collaborative filtering or content-based filtering to acquire a Top-N recommendation [1, 7, 9], while others focused on a bottom-up approach to first analyze datasets individually [3]. Some other papers even took a novel hybrid approach for creating a lightweight and fast recommendation algorithm for jobs that met the goals of scalability, incremental model updates, and fast score calculations for jobs by combining a content-based approach with KNN. [4]. Not all recommender systems went with a collaborative filtering approach. More specifically, some researchers found that collaborative filtering won't be enough to create a useful job recommendation system due to user inactivity, so they proposed to use collaborative filtering combined with a content-based approach after discovering user inactivity in major datasets [5].

Instead of traditional collaborative filtering or content-based approaches, there was also research that combines temporal learning with sequence modeling to capture complex user-item activity patterns to improve job recommendations, and further optimized them using recurrent neural networks [6]. Gradient Boosting Decision Trees have also been experimented with to predict positive implicit interactions with job offers [7]. The Hawkes Process was the winning submission for the RecSys Challenge 2016, known for its novel approach of generating recommendations by modelling temporal activity patterns using a self-exciting point process [9].

One major problem that many collaborative filtering recommender systems face is the cold-start setting. Cold-start refers to when a recommender system fails to provide accurate or meaningful recommendations to users when using collaborative filtering. One of the main causes of cold-start is simply because there is either a new user, or item in the system. Using Amazon as an example, cold-start can occur when 1) a new user is created, or 2) a new item is added to Amazon's catalog. Both scenarios, a new user or item, will cause the recommender system to fail to attain meaningful recommendations simply due to a lack of data. Once a user starts using their new account, or an item starts getting some attention/popularity, collaborative filtering will perform better since implicit data will get tracked. Many research papers, specifically from the RecSys Challenge 2017, have found some solutions to the cold start problem regarding job recommendation systems [10, 11, 12, 13, 14]. Researchers used more explicit data in the users' and jobs' profiles instead of relying on implicit data [10], since implicit data is usually empty upon fresh user or job profile creations. The most common solution to the cold-start problem is to include a content-based approach in the recommender system, instead of solely using collaborative filtering [10, 12, 14].

Most of this past research had the goal of optimizing the resulting job recommendations to accurately match a user's interests, and they don't optimize the overall state of the job market. Although our recommender system is driven by the DAA rather than collaborative or content-based filtering, we still can use some of this previous work to further optimize our recommender system. In this paper, we will also make use of some collaborative filtering to discover new preferences that will be fed in as input into our DAA-driven recommender system in order to maximize job market stability.

## 2.2    Present-Day Applications of DAA

In the economics field, many researchers have seen the DAA in use for finding matches for college admissions, dating websites, and even for employment in the medical field. The DAA is mainly used for any two-sided matching problem. There are one-to-one matching problems, such as the famous stable marriage problem which the original Gale-Shapley DAA solves. The incentive to use the DAA for one-to-one matching is for its stability, because stability guarantees Pareto efficient matching and individually rational matching [15]. The DAA has also been used to solve some many-to-one matching problems, such as college admissions [15]. The college admission problem is similar to the stable marriage problem but differs in the sense that a college can have multiple students whereas a man should only have one woman (in most societies). Even though the DAA was originally created to solve two-sided matching problems, there has been some use for it regarding one-sided matching, such as the house allocation problem [15].

For our recommender system, we will be mainly focusing on the classic one-to-one matching aspect of the standard two-sided matching problem. While the DAA offers stable one-to-one matches, we want to extend the base version of the DAA to support multiple sets of one-to-one matches to acquire multiple recommendations rather than just one. We want to ensure that our recommendations are both Pareto efficient (stable), and individually rational, which is what the standard one-to-one DAA offers.

In the computer science field, the DAA hasn't really gained too much attention yet. The Gale-Shapley deferred acceptance algorithm is mainly known for solving the stable marriage problem. However, Hosam AlHakami, Feng Chen and Helge Janicke found a new use for the DAA that's different from the typical matching market application. These researchers were able to extend the DAA to create a new algorithm for detecting cloned code in order to preserve intellectual property and prevent negative effects in industry software [16]. They used the DAA to detect code clones with high scalability and accuracy by treating code fragments as a two-sided matching problem.

Like the code clone detection algorithm, we want to take advantage of what the DAA has to offer and use it as an approach to stabilize the job market through the use of a recommender system with novel machine learning optimizations. With the DAA offering Pareto efficient and individually rational matches through match stability, we can guarantee that our recommender system will reach our desired accuracy and help minimize vacancy in the job market.

# Chapter 3

# Multi-Match Deferred Acceptance Algorithms (MMDAA's)

## 3.1 The DAA

Figure 3.1 illustrates the DAA's input and output. The input of the DAA is any equal number of candidates and employers along with their preferences of each other. The top two tables of Figure 3.1 show the preference data of each candidate and employer. These preference tables are in descending order, from most preferred to least preferred. For example, candidate 1 prefers employer 2 first, employer 3 second, and employer 1 third. The output of the DAA is a *stable* one-to-one matching between the candidates and employers. As stated earlier, the term *stable*, refers to that these matches guarantee that there are no two people, a candidate and an employer, who would both rather have each other than their current partners that resulted from the DAA. In other words, no match exists that would make a candidate and an employer more fit for each other. An example of a stable one-to-one matching output can be found in Figure 3.1.

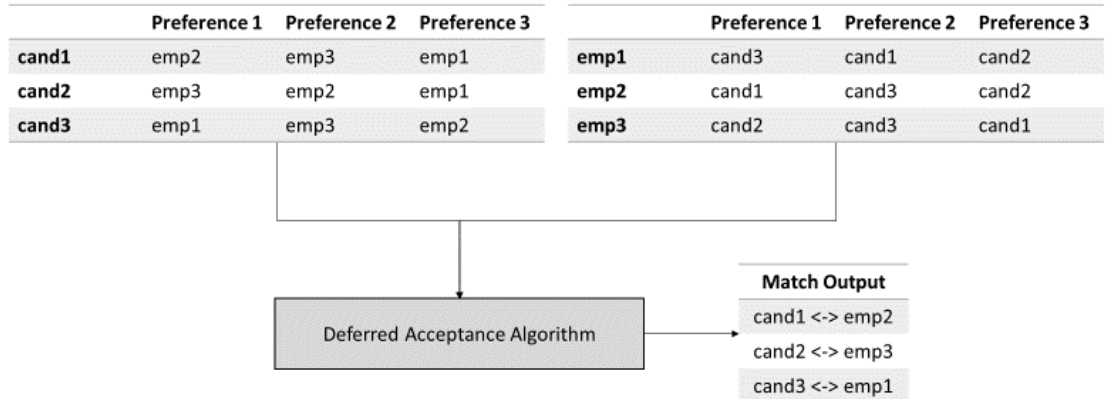

Figure 3.1: DAA Input & Output

The DAA is a simple greedy algorithm. The first step of the DAA is to initialize all candidates *c* and all employers *e* as "free" [Figure 3.2, Line 2]. In other words, we set everyone's match status to "free." Then, we go through every free candidate *c* and make proposals to *c*'s most preferred employer *e* [Figure 3.2, Line 4].

There are two things that can happen once *c* has proposed to *e*. The first case is simple; *e* is free, so we simply "engage" *c* and *e* together [Figure 3.2, Lines 5-6]. The second case is a bit more complicated; *e* is not free because *e* was engaged already to some other candidate *c'* [Figure 3.2, Line 7]. In this situation, the DAA must see who *e* prefers more, the candidate of the current engagement *c'*, or the new proposer *c*. If *e* prefers candidate *c'* more, then we will continue to let *c'* and *e* be engaged. If *e* prefers candidate *c* more, then we will break off the existing engagement between *c'* and *e*, and create an engagement between *c* and *e* [Figure 3.2, Lines 8-12]. We repeat this process until there are no more "free" candidates [Figure 3.2, Line 3]. In other words, once every candidate *c* has an engagement, we can end the DAA and return all engaged pairs (*c*, *e*) [Figure 3.2, Line 16]. Note that the original DAA applies to equal sized sets of data (*n* candidates and *n* employers). McVitie & Wilson adapted the DAA to support unequal sized sets of data [21]. Lines 13-15 in the pseudocode below allow the DAA to find stable matches for unequal sets. This is achieved by setting a maximum "count" on the number of proposes made.

```
1.   function DAA() {
2.       Initialize all c ∈ C and e ∈ E to free
3.       while ∃ free candidate c who still has a employer e to propose to {
4.           e = first employer on c's list to whom c has not yet proposed
5.           if e is free
6.               (c, e) become engaged
7.           else some pair (c', e) already exists
8.               if e prefers c to c'
9.                   c' becomes free
10.                  (c, e) become engaged
11.              else
12.                  (c', e) remain engaged
13.              if proposalCount > maxNumOfProposals
14.                  break
15.              increment proposalCount
16.      }
17.      return all engaged pairs(c, e)
18.  }
```

Figure 3.2: DAA Pseudocode

The runtime complexity of the DAA is $O(n^2)$, where $n$ is the number of employers or candidates. This is because we have an outer while loop that at the very minimum traverses through each of the $n$ candidates. And inside of this while loop, we must linearly traverse through each candidates' preference list of the employers, which is also of size $n$ due to each candidate having a preference on each employer. Hence, two nested loops of size $n$ yield a runtime complexity of $O(n^2)$.

The drawback of this algorithm is that it requires the preferences datasets to be equally sized. $n$ candidates with $n$ preferences of each employer, and vice versa. If the preference datasets aren't square and dense ($n$ by $n$), then DAA may fail to find a stable one-to-one matching for everyone in both parties. In other words, there may be some candidate or employer left without a match at the end of the DAA. Because of this tough constraint, the DAA is hard to use in real-world applications.

To use the DAA in the real-world application of matching candidates with employers, we will need to have square and dense preference datasets. To have square and dense preference datasets, we would have to ensure that 1) every single candidate has a preference on every employer, and every single employer has a preference on every candidate; 2) the number of candidates is equal to the number of employers. Both requirements are highly unlikely to be met in the real world, since there are usually more candidates than jobs, and gathering the preference data on every single person will cause the DAA to not be scalable.

## 3.2    The Normal MMDAA

Since the DAA fails to find a stable one-to-one matching when the preference datasets aren't square and dense, we propose the Normal Multi-Match Deferred Acceptance Algorithm (Normal MMDAA). The Normal MMDAA liberates the strict requirements of the DAA, while still providing stable matches. It offers a more scalable solution for finding stable matches since it allows the preference datasets to be non-square and sparse. Going back to an earlier example, if we are trying to match job seekers (candidates) to jobs (employers), we could use the DAA to find a stable set of matches if the preference datasets were square and dense. Since these preference datasets are not likely to be square and dense, the MMDAA is an alternative solution.

The Normal MMDAA architecture is shown in Figure 3.3, it requires two things as input, 1) the candidates' preferences of each employer, and 2) the employers' preferences of each candidate. The input to the Normal MMDAA is the same as the regular DAA. The key difference between these two algorithms is the output. The DAA outputs a *single* set of stable matches that are one-to-one, while the Normal MMDAA outputs *multiple* sets of stable matches that are one-to-one. The main reason why we want the Normal MMDAA to output multiple sets of stable matches is to ensure that everyone (candidates and employers), will at least receive one match, as compared to receiving no match from the regular DAA.

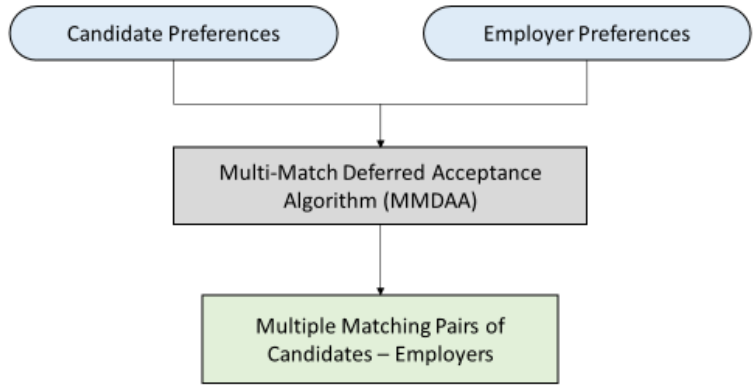

Figure 3.3: Normal MMDAA Architecture

Figure 3.4 depicts a more detailed example of the Normal MMDAA. The input of the Normal MMDAA is several candidates and employers along with their preferences of each other. In Figure 3.4, the top two tables show the preference data of each candidate and employer. Again, these preference tables are in descending order, from most preferred to least preferred. The output of the Normal MMDAA is *multiple* sets of *stable* matches between the candidates and employers. These matches from the Normal MMDAA are also in descending order. Taking the match results for candidate 1 as an example, the first (and best) match is employer 2, the second best match is employer 1, and the third (and worst) match is employer 3. Note that here, even though candidate 2 was forced to receive no second match (N/A means no match), candidate 2 was given a match in a later round. Candidate 3 received only one match, because candidate 3 only had a single preference. The key idea to take away from the Normal MMDAA is that someone can receive no match for a specific round (N/A), but continue to obtain less important matches. In the regular DAA, if someone was forced to have no match, then he/she would just remain unmatched since there is only a single set of matches.

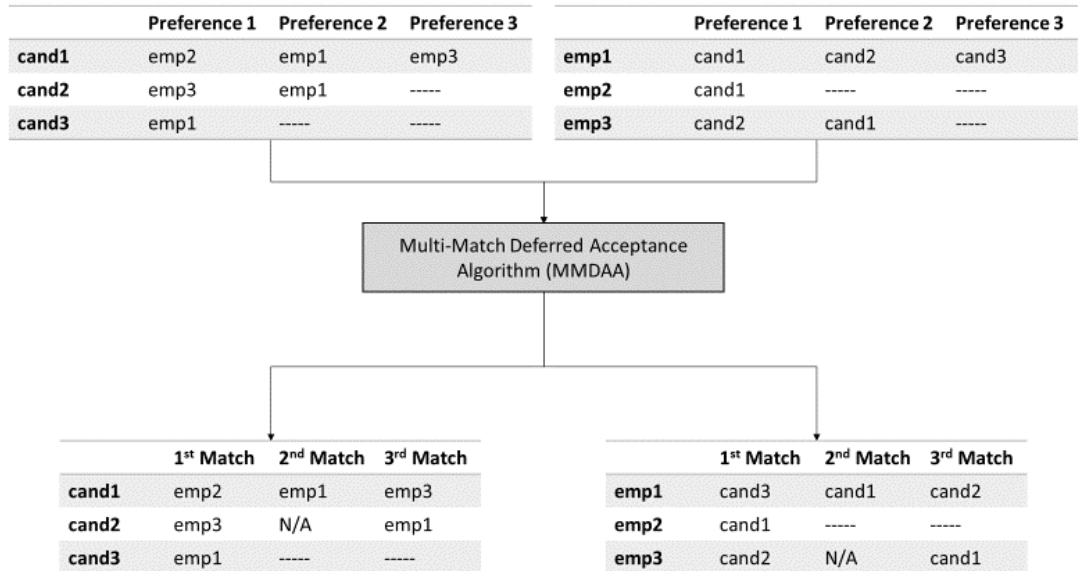

Figure 3.4: Normal MMDAA Input & Output

The Normal MMDAA uses the DAA multiple times, or rounds, to obtain multiple different sets of stable one-to-one matches. In other words, we basically run the DAA on the preference datasets, and obtain a single set of stable one-to-one matches. Then we remove these stable matches from the original preference datasets and run the DAA again. We remove the matches from the preference datasets to ensure we won't get the same, optimal matches again. We repeat this process of running the DAA and altering the preference datasets until we run out of preference, or until we reach our desired number of *stable* matches. Figure 3.5 shows an example of a single iteration of the Normal MMDAA.

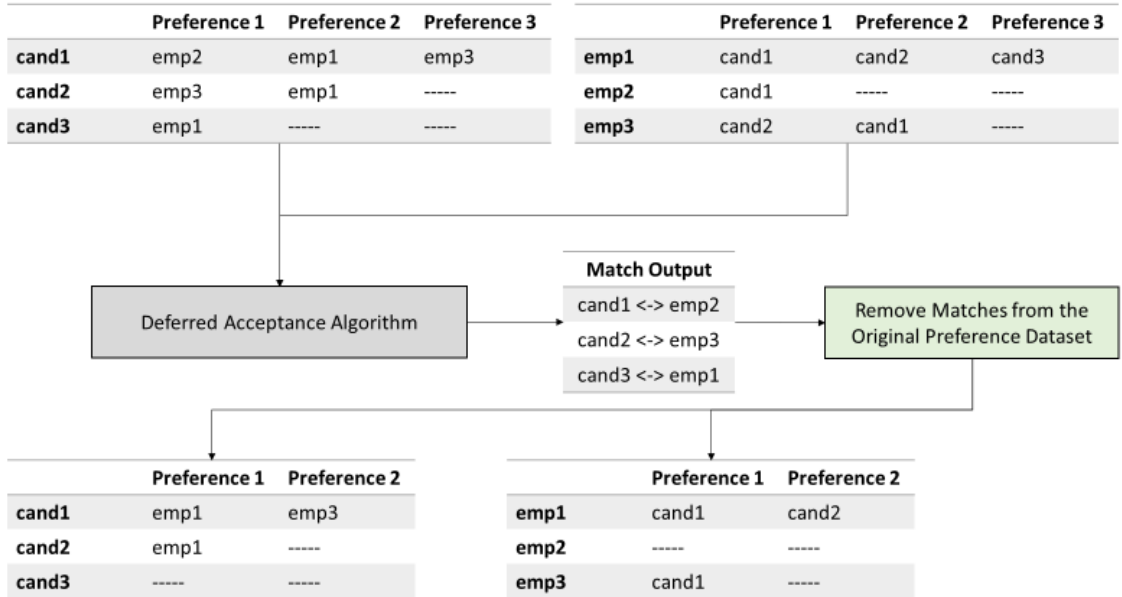

Figure 3.5: MMDAA, Single Iteration Example

The Normal MMDAA is an iterative algorithm that makes use of the greedy DAA. Figure 3.6 contains the pseudocode for the Normal MMDAA. As stated earlier, Normal MMDAA uses the DAA to obtain a single set of stable one-to-one matches [Line 6]. After finding a set of stable matches, we then go through the preference datasets and remove any preferences found in this set of stable matches [Lines 7-8]. After altering the preference datasets, we append this set of stable matches to *allMatches*, and increment the total number of matches obtained [Lines 9-10]. We repeat this process until the Normal MMDAA meets a termination case [Line 5]. The first termination case is if the preference datasets are empty, which is caused from removing matches from the preference datasets up to the point that they are empty. This commonly occurs when 1) there are small preference datasets, or 2) there is a high number of desired matches. The second termination case is simply if we met the desired number of matches. In our case, we usually desire just 5 matches (the top 5 matches). Once the Normal MMDAA has hit a termination case, we return the list of all obtained stable one-to-one matches [Line 12].

```
1.   function MultiMatchDAA {
2.       numberOfMatches = k // The maximum number of matches to find.
3.       matchNumber = 0 // A counter for the current number of matches.
4.       allMatches // A List to store all matches.
5.       while preference datasets are not empty or matchNumber >= numberOfMatches {
6.           matches = DAA()
7.           for every match in matches
8.               remove match from original preference datasets.
9.           append matches to allMatches
10.          matchNumber++
11.      }
12.      return allMatches
13.  }
```

Figure 3.6: MMDAA Pseudocode

The runtime complexity of the Normal MMDAA is $O(kn^2)$, where $n$ is the number of candidates, and $k$ is the number of desired matches. The while loop of the Normal MMDAA will run at most $k$ iterations. Inside of this while loop, we must find a stable set of one-to-one matches by using the DAA, which has a runtime of $O(n^2)$. Hence, two nested loops of size $k$, and $n$, respectively, yield a runtime complexity of $O(kn^2)$. Typically, $k$ is a small value, so the runtime complexity of the MMDAA can be treated as $O(n^2)$.

Similar to the DAA, the Normal MMDAA also has some drawbacks. While the Normal MMDAA doesn't require the preferences datasets to be square and dense, it doesn't guarantee that every candidate or employer will get at least one match. The Normal MMDAA does indeed give a higher chance for everyone to find at least one match, but it ends up facing the same limitation as the regular DAA, which is obtaining no matches due to sparse datasets. The Normal MMDAA is built on top of the regular DAA, so it also inherits some of its limitations. The square and dense datasets limitation was partly avoided in the Normal MMDAA, but not completely solved.

Referring back to Figure 3.4, we can see that sometimes a candidate or employer will be forced to be unmatched (N/A) for a specific matching round, but continue to receive matches in later rounds. The Normal MMDAA is useful in the sense that all candidates or employers will have a high chance of receiving at least one match, because the Normal MMDAA gives multiple opportunities for everyone to get a match (through multiple DAA matching rounds). However, there are still two major limitations with the Normal MMDAA.

The first limitation of the Normal MMDAA is simply what if a candidate or employer remains unmatched for all $k$ ($k$ is the desired number of matches) rounds. If this occurs, then that candidate or employer will simply receive no match at all from the Normal MMDAA. We have ran some experiments, and found that some candidates and employers were actually forced to remain unmatched for more than 7 continuous rounds. As stated earlier, $k$ is typically some small value, which means this limitation can occur quite commonly. One possible solution would be to increase $k$ to some large value. However, this wouldn't really benefit the candidate or employer that was forced to remain unmatched for continuous rounds, because his or her first assigned match will be of low importance. (For example, a candidate sitting out of 10 rounds in a row, and finally getting a match in the 11[th] round. How important is this 11[th] match, if someone else was a better match for everyone one of this candidate's preferences in the previous 10 rounds?).

The second limitation of the Normal MMDAA is about the importance of the matches assigned to candidates and employers after they were forced to remain unmatched during a specific matching round. Going back to the example in Figure 3.4, candidate 2 was forced to remain unmatched in the 2$^{nd}$ matching round. This was because candidate 2's only preference was employer 1, but employer 1 was better matched with candidate 1 according to their preferences. During the 3$^{rd}$ round, candidate 2 finally got to match with employer 1, since there weren't any other preferences. Even though candidate 2 was able to match with employer 1 in a later round, the importance of this match is quite low, since employer 1 was better fit for candidate 1 in an earlier round. We also can't just "shift" the match results to the left to cover up the no matches, because this would violate the one-to-one stability rule of the DAA. In other words, we can't make the 3$^{rd}$ match be the 2$^{nd}$ match for candidate 2, because then employer 1 would have matched with more than one candidate in a single round, violating the *stability* of the matches. Overall, these candidates or employers who receive matches after being forced to be unmatched in a round, can still be considered to have no matches, because these later matches are of little value.

Both of these limitations affect the application of the Normal MMDAA to real-world problems. Even though the Normal MMDAA is more scalable than the DAA, it still has its own limitations that are yet to be solved. Part of the reason these limitations occur is still because of the non-square and sparse preference datasets, since this is what causes candidates or employers to be unmatched. On top of the non-square and sparse preference datasets, the Normal MMDAA is actually altering the sparsity and "squareness" of the preference datasets as it progresses through each matching round, because it's removing each round's set of stable matches from the preference datasets. Therefore, the root limitation of the Normal MMDAA is still this restriction of square and dense preference datasets.

## 3.3   The LMF-MMDAA

The Normal MMDAA helps loosen the requirements for the DAA, but it still faces its own limitations. As stated earlier, the source of these limitations is still due to the shape and sparsity of the preference datasets. To truly offer meaningful, stable one-to-one matches with the Normal MMDAA or DAA, we must have dense preference datasets. To achieve this, we propose to perform Low-Rank Matrix Factorization (LMF), or Non-Negative Matrix Factorization (NNMF) [20], on the preference datasets. LMF achieves two new advantages, 1) preference datasets are guaranteed to be dense, and 2) we infer new preferences on employers and candidates through collaborative filtering.

Collaborative filtering is highly used in recommender systems, such as Netflix. The idea behind collaborative filtering with preference datasets is that we predict new preferences based off users who share common preferences. In other words, users are indirectly collaborating with each other to find a common understanding of what preference best aligns with a certain candidate or employer. LMF allows us to find new preferences on employers and candidates who never actually met each other, based off candidates and employers who share common preferences/interests. With these LMF'd preferences, we can find *meaningful* matches between candidates and employers who never met or had a preference on. The key idea here is that LMF will provide us with preference datasets that contains a preference ranking for every candidate and employer in the system (dense datasets), through collaborative filtering.

Figure 3.7 shows the LMF-MMDAA architecture. As usual, it still requires two things as input, 1) the candidates' preferences of each employer, and 2) the employers' preferences of each candidate. Rather than throwing the original preference datasets as input directly into the Normal MMDAA, we first want to perform LMF on them. After computing our LMF'd preference datasets, we pass these new dense preferences into the Normal MMDAA. This process will be known as LMF-MMDAA. LMF-MMDAA still outputs *multiple* sets of stable matches that are one-to-one, and it offers a lower rate of unmatched candidates and employers per round, as compared to the Normal MMDAA with the original preferences.

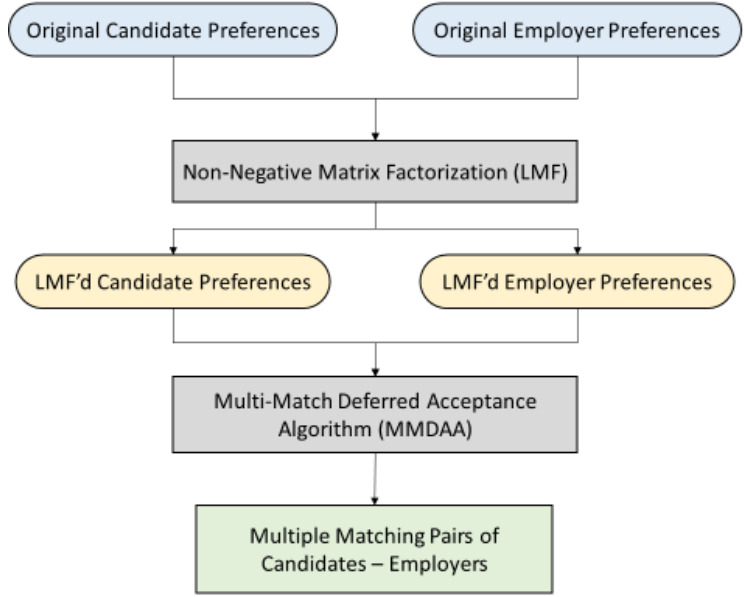

Figure 3.7: LMF-MMDAA Architecture

A more detailed example of the LMF-MMDAA can be found in Figure 3.8. With LMF-MMDAA, we want to convert our sparse preference datasets into dense preference datasets before running the MMDAA. The first two tables represent the original preferences of the candidates and employers. These preferences are the same ones used in the original Normal MMDAA example (Figure 3.4). The LMF'd preferences can be found in the middle two tables of Figure 3.8. These LMF'd preferences, compared to the original preferences, have all its columns filled. In other words, every candidate has a preference ranking on every employer, and vice versa. This is the key feature of LMF; it allows us to satisfy the strict requirements of the DAA algorithm without actually having to ask each candidate to manually choose a preference rank on every single employer, and vice versa. LMF allows the MMDAA to be applied to real-world problems because we can densify the preference datasets ourselves. Continuing with the example, we then feed the LMF'd preferences into the MMDAA, and receive multiple *stable* matches as output.

In the Normal MMDAA (Figure 3.4), candidate 2 was forced to receive no match in the second round, but was given a less important match in the third round. Also in the Normal MMDAA (Figure 3.4), candidate 2 only received two matches and candidate 3 only received one match. This was simply because they only had a small number preferences. As shown in Figure 3.8, the LMF-MMDAA outputs matches for every candidate and employer for every round. In other words, everyone had obtained 3 different matches.

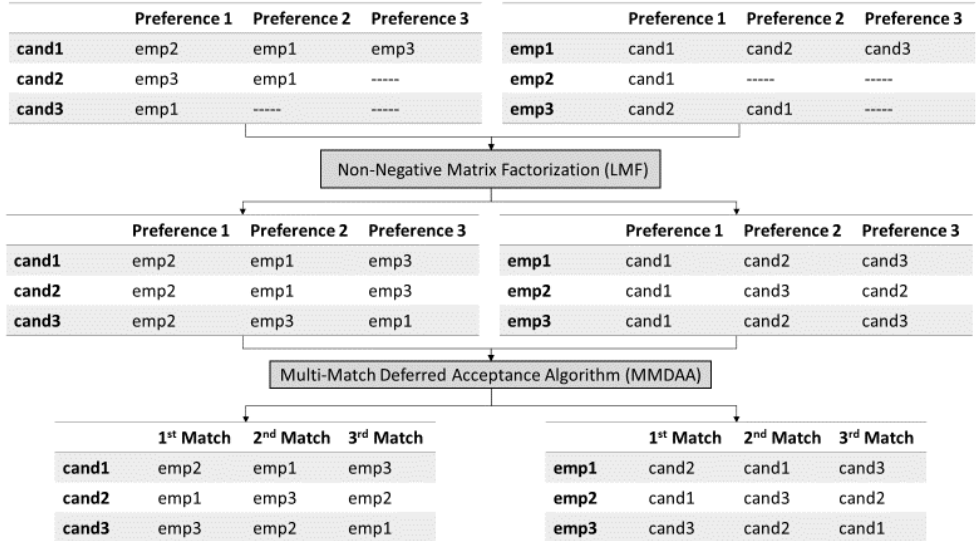

Figure 3.8: LMF-MMDAA Input & Output

MMDAA's objective is to ultimately minimize the number of unmatched candidates and employers. As stated in the previous section, Normal MMDAA still faced its own limitations, 1) unmatched candidates and employers for continuous matches, and 2) the matches assigned in later rounds to those who were previously unmatched have little importance. Both of these limitations were again caused by the source requirement of the regular DAA – preference datasets must be square and dense. Since LMF forces the preference datasets to fulfill the dense requirement of the DAA, the limitations of the Normal MMDAA were fixed. LMF-MMDAA *almost* guarantees that every candidate and employer will receive a match in every round. The reason we use the term "almost", is because LMF-MMDAA doesn't account for the sparsity changes as the MMDAA progresses through each round. As mentioned earlier, when the MMDAA moves to the next round, it removes the stable matches of that current round from the preference datasets. Because of this, the sparsity of the preference datasets will change. LMF-MMDAA guarantees that the preference datasets will be dense when we *first start* the matching algorithm, but it doesn't guarantee that the preference datasets will remain dense as the algorithm progresses. However, the LMF-MMDAA still shows a great improvement from the Normal MMDAA in terms of the number of unmatched candidates and employers.

The main issue with LMF-MMDAA isn't regarding the number of unmatched candidates and employers, it's more on the actual accuracy of the matches. The most challenging part of the LMF-MMDAA algorithm is filling the missing rankings with accurate values, i.e., the jobs ranked high for a candidate are indeed relevant to the candidate. The Low-rank Matrix Factorization (LMF) technique is applied to the preference input matrices. LMF produces two factor matrices whose product results in a dense matrix with decimal values at each entry. The factor matrices are "learned" by minimizing the difference between the explicit rankings and the value obtained at that entry by their multiplication. This preserves the given rankings, while deriving optimal values for the missing ones based on the rankings of all the candidates/employers. In order to convert the decimal matrix entries to discrete ranks, we iterate over each generated ranking list and scale-up the decimals to integers such that the relative order between ranks is preserved. However, it is not guaranteed that the resulting order follows the originally stated rankings---the "learning" is not exact. Several approaches are possible. We can use the LMF output as is. In this case, the stated rankings are taken into account only to the extent they are preserved by LMF. We can preserve the relative ordering between the given ranks. This stops LMF to change the stated order, however, the missing rankings can interleave with the given ones. The last alternative is to fully keep the given rankings and only use the LMF output for the ordering of the non-stated entries. We experimented with each of these solutions and found empirically that preserving only the relative order of the given ranks generates the best results. Thus, we use this approach in LMF-MMDAA. The computation of the factor matrices, i.e., the LMF training, is a time-consuming process. However, this is a pre-processing step that is done once and offline in LMF-MMDAA. Since the online step consists of performing MMDAA on the dense output produced by LMF, the algorithmic complexity of MMDAA is preserved.

Since we performed LMF on the original preference datasets, the old original preferences were overwritten with collaborative filtered preferences. Because of this, we don't know how "good" or "accurate" these new preferences are. To ensure that the match results are still accurate with LMF'd preferences, we created a metric called *Displacement*. *Displacement* essentially measures how successful the output of an MMDAA algorithm really is. We measure this success by relating the output matches with how close they were to the matching algorithm's input preferences. The steps for calculating *displacement* are: For each user, we take a match from a given round and see where it is in the user's preference list. Figure 3.9 shows an example of how to calculate the displacement based off the input/output from Figure 3.8. To calculate a displacement value for a specific candidate on a certain round, we look at the match value and see where it's located at in the preference list. For example, candidate 1's displacement on round 1 is 0 because candidate 1's first match was employer 2, which was the $1^{st}$ preference. We consider the first preference as a displacement value of 0 (start counting from 0). Another example, candidate 2's displacement on round 2 is 2 because candidate 2's second match was employer 3, which was the third preference, hence this is a displacement of 2 (start counting from 0). Summing all the candidates' displacement for a round will yield the total displacement. Average displacement is obtained by dividing the total displacement by the number of candidates. We will use the average displacement as a measure for match accuracy. The lower the average displacement, the more accurate the matches are with respect to the input preferences.

|  | Preference 1 | Preference 2 | Preference 3 |  | $1^{st}$ Match | $2^{nd}$ Match | $3^{rd}$ Match |
|---|---|---|---|---|---|---|---|
| cand1 | emp2 | emp1 | emp3 | cand1 | emp2 | emp1 | emp3 |
| cand2 | emp2 | emp1 | emp3 | cand2 | emp1 | emp3 | emp2 |
| cand3 | emp2 | emp3 | emp1 | cand3 | emp3 | emp2 | emp1 |

|  | Round 1 | Round 2 | Round 3 |
|---|---|---|---|
| cand1 Disp. | 0 | 1 | 2 |
| cand2 Disp. | 1 | 2 | 0 |
| cand3 Disp | 1 | 0 | 2 |
| Total Disp. | 2 | 3 | 4 |
| Avg Disp. | 2/3 | 3/3 | 4/3 |

Figure 3.9: Displacement Metric Calculation for LMF-MMDAA

After a few experiments, we found that LMF-MMDAA has a higher displacement (worse match accuracy) than the Normal MMDAA. LMF-MMDAA gains a significantly lower rate of unmatched candidates and employers at the cost of match accuracy. There is a trade-off between match accuracy and the number of unmatched users. We will focus on why LMF-MMDAA has a worse match accuracy later in the experiments/results section. LMF-MMDAA allows true use of this matching algorithm in real-world applications, because it guarantees that the preference datasets will always be dense. However, the match results aren't as optimal as they can be, and it is still possible to have an occasional unmatched user due to how the sparsity of the preference datasets changes as the matching algorithm progresses to different rounds.

## 3.4    The Mixed MMDAA

LMF-MMDAA solved the problem having a high number of unmatched candidates and employers, but at the cost of worse match accuracy, or *displacement*. Our goal is to optimize the tradeoff we face when running the MMDAA. In other words, we want to minimize the number of unmatched candidates and employers, and also maximize the match accuracy. To achieve this, we propose to use a Mixed Multi-Match Deferred Acceptance Algorithm, or Mixed MMDAA.

The architecture of Mixed MMDAA can be found in Figure 3.10. As input, this algorithm requires matches rather than preferences. The Mixed MMDAA requires 4 sets of matches: 1) The candidate match results from the Normal MMDAA, 2) The employer match results from the Normal MMDAA, 3) The candidate match results from the LMF-MMDAA, and 4) The employer match results from the LMF-MMDAA. This algorithm won't necessarily be running a matching algorithm, but instead it simply combines different match results. Before running the Mixed MMDAA, the Normal MMDAA and the LMF-MMDAA must first be ran, because we need to acquire the different match results in order to combine them. The Mixed MMDAA still outputs *multiple* sets of stable matches that are one-to-one, and it offers the lowest rate of unmatched candidates and employers per round with better *displacement* than the matches outputted by LMF-MMDAA.

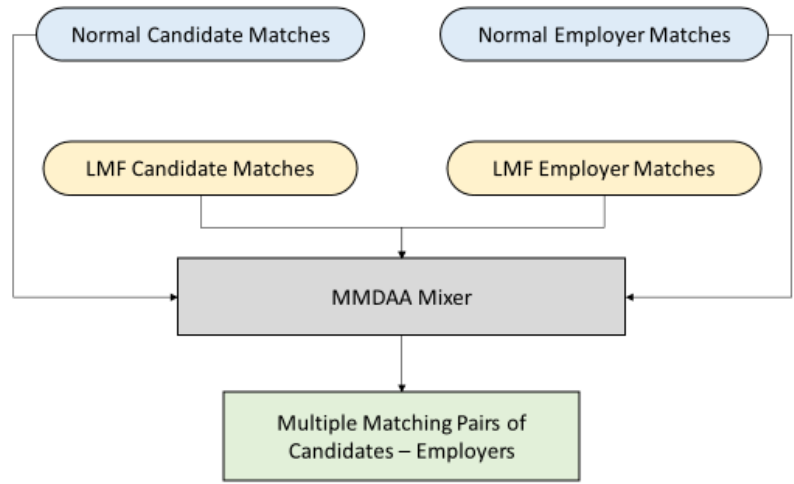

Figure 3.10: Mixed MMDAA Architecture

Figure 3.11 shows an example of how the Mixed MMDAA works. The first two tables represent the Normal MMDAA matches of the candidates and employers, and the middle two tables represent the LMF-MMDDA matches of the candidates and employers. For simplicity, these matches are the matches we acquired in the previous examples, (Figure 3.4 and Figure 3.8). Continuing with the example, we then feed these matches into the Mixed MMDAA, and as output we receive multiple *stable* matches that optimize the number of unmatches and match accuracy.

Looking more closely at the input of Figure 3.11, we can see that the match results of the Normal MMDAA forced candidate 2 receive no match in the second round, but was given a less important match in the third round. Again, this later match is considered less important because someone else (in this case candidate 1), had a better match with whoever candidate 2 was supposed to match with originally (employer 1). Also note that in the Normal MMDAA, not all candidates received 3 matches. This was simply because they only had a small number of preferences (refer to Figure 3.4). As for LMF-MMDAA matches, we have 3 matches for all candidates, but at the cost of a higher *displacement*. The goal here would be to combine the different match results into one more optimal match result. The Normal MMDAA offers a low *displacement*, at the cost of a high rate of unmatches, while the LMF-MMDAA offers a low rate of unmatches, at the cost of a high *displacement*.

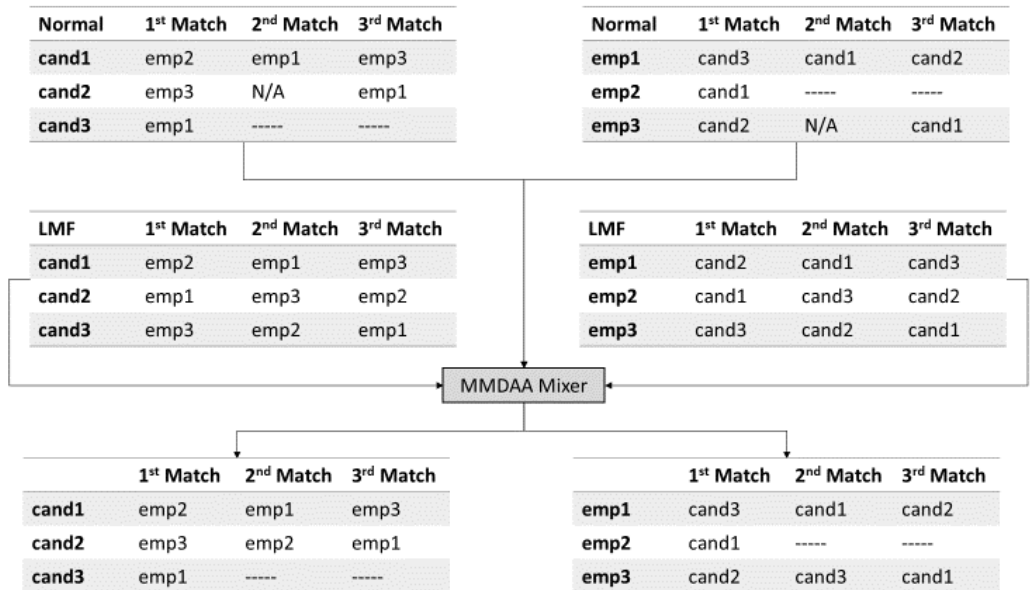

Figure 3.11: Mixed MMDAA Input & Output

To measure how efficient the Mixed MMDAA's output really is, we need to measure the number of unmatches and the *displacement*. We can see from the output matches in Figure 3.11, that there are no unmatched candidates and employers. The Mixed MMDAA preserves or improves the number of unmatched candidates and employers from the LMF-MMDAA. At the same, it offers a more efficient *displacement* compared to the LMF-MMDAA. Figure 3.12 shows the *displacement* calculation of the Mixed MMDAA. Comparing this to the *displacement* calculation of the LMF-MMDAA in Figure 3.9, we can see that the Mixed MMDAA has a lower average *displacement* for every round.

As stated earlier, we measure *displacement* by relating the output matches with how close they were to the matching algorithm's input preferences. Note that since we are combining 2 different matching results into a single matching result, we have to compare each match with the correct input preference dataset. In other words, if the match we are calculating displacement for is from the normal preferences then we calculate displacement based off the normal preference dataset, and vice versa for LMF matches/preferences. The *displacement* calculation of the above example can be found in Figure 3.12. For example, candidate 1's displacement on round 2 is 1 because candidate 1's second match was employer 1, which was the 2nd preference of the normal preference dataset. Candidate 2's displacement on round 2 is 0 because candidate 2's second match was employer 2, which was the 1st preference of the LMF preference dataset. Since candidate 3 had no more preferences on round 2, it didn't receive a match. Keep in mind that since candidate 3 won't be counted in the average *displacement* for round 2, since candidate 3 didn't need to participate in round 2.

| (Original) | Preference 1 | Preference 2 | Preference 3 | | 1st Match | 2nd Match | 3rd Match |
|---|---|---|---|---|---|---|---|
| cand1 | emp2 | emp1 | emp3 | cand1 | emp2 | emp1 | emp3 |
| cand2 | emp3 | emp1 | ----- | cand2 | emp3 | emp2 | emp1 |
| cand3 | emp1 | ----- | ----- | cand3 | emp1 | ----- | ----- |

| | | | | | Round 1 | Round 2 | Round 3 |
|---|---|---|---|---|---|---|---|
| (LMF) | Preference 1 | Preference 2 | Preference 3 | cand1 Disp. | 0 | 1 | 2 |
| cand1 | emp2 | emp1 | emp3 | cand2 Disp. | 0 | 0 | 1 |
| cand2 | emp2 | emp1 | emp3 | cand3 Disp | 0 | ----- | ----- |
| cand3 | emp2 | emp3 | emp1 | Total Disp. | 0 | 1 | 3 |
| | | | | Avg Disp. | 0 | 1/2 | 3/2 |

Figure 3.12: Displacement Metric Calculation for Mixed MMDAA

The Mixed MMDAA is a simple algorithm that combines two matching outputs. The Mixed MMDAA requires the normal matches and LMF matches of each candidate and employer [Figure 3.13, Line 10]. Then for each candidate and employer, we need to find all of the matching rounds resulted in no match, and fill them in with a substitute match [Figure 3.13, Lines 11-14]. The no match filling process is handled by the fillNoMatches function described below. The fillNoMatches function takes two things as input, the normal matches of a user, and the LMF matches of a user [Figure 3.13, Line 1]. The Mixed MMDAA iterates over all the normal matches and checks if there is a match for each round [Figure 3.13, Lines 2-3]. If there is a round that resulted in no match (denoted by -1), then we need to "fill" this no match with the next best match found in the LMF matches [Figure 3.13, Line 3-4]. We "fill" no matches by going through the LMF matches and finding a match that satisfies two requirements: 1) this new match must not have been matched to this particular user in any other round, otherwise this user will have duplicate matches, and 2) this new match must not have been matched to any other users, otherwise two users would share the same match for a single round which violates the *stability* and the *one-to-one* features of these matches. After filling all the rounds that have no matches, we simply return the new set of matches [Figure 3.13, Line 7].

```
1.  function fillNoMatches(normalMatches, lmfMatches) {
2.      for i = 0 to normalMatches.size {
3.          if (normalMatches[i] == -1) {
4.              normalMatches[i] = findNextBestMatch(normalMatches, lmfMatches)
5.          }
6.      }
7.      return normalMatches
8.  }
9.
10. function MixedMMDAA(candMatches, lmfCandMatches, empMatches, lmfEmpMatches) {
11.     for i = 0 to candidates.size
12.         fillNoMatches(candMatches[i], lmfCandMatches[i])
13.     for i = 0 to employers.size
14.         fillNoMatches(empMatches[i], lmfEmpMatches[i])
15. }
```

Figure 3.13: Mixed MMDAA Pseudocode

The runtime complexity of the Mixed MMDAA is $O(n * m * k)$, where $n$ is the number of employers or candidates (the larger of the two), $m$ is the number of matching rounds from the Normal MMDAA, and $k$ is the number of matching rounds from the LMF-MMDAA. The runtime complexity is $O(n * m * k)$, because we have an outer for loop that traverses through each of the $n$ candidates and employers (inside the MixedMMDAA function). And inside of this for loop, we must linearly traverse through each match, which is of size $m$. When traversing through each match, if there is a round that resulted in a no match, then we will have to fill it. Filling this match requires us to traverse through the LMF-MMDAA Matches, which is of size $k$. Hence, three nested loops of size $n$, $m$, and $k$ yields a runtime complexity of $O(n * m * k)$, where $m$ and $k$ should be $<< n$.

The Mixed MMDAA optimizes the trade-off between match accuracy and the number of unmatched users by using the Normal MMDAA match results as much as possible, and using the LMF-MMDAA results only when there are rounds with no matches. This allows us to maintain the low rate of unmatched users while maintaining an effective *displacement* similar to the *displacement* of the Normal MMDAA.

# Chapter 4

# Analysis on the Various MMDAA's

## 4.1  Metrics and Measurement Definitions

To evaluate how well each of the three matching algorithms perform (Normal MMDAA, LMF-MMDAA, and Mixed MMDAA), we measure two attributes. The first measurement used for our evaluation is *displacement*. Displacement, as explained in detail in Chapter 3.3, measures the success, or match accuracy, of a particular set of matches. As demonstrated earlier, displacement is calculated by relating the output matches with how close they were to the matching algorithm's input preferences. The lower the average displacement, the more accurate the matches are with respect to the input preferences. Average displacement is defined as the average displacement across all candidates, or employers, for a given round. Note that only the candidates, or employers, that receive a match for a given round will be included in that round's average displacement. In other words, if a candidate, or employer, happened to receive no match, they will not be included in the average displacement calculation.

Our second measurement used to evaluate our matching algorithms are *withholdings*. Withholdings measures the number of candidates, or employers, that were required to be withheld from receiving a match for a particular round. In other words, withholdings measure the number of matchless candidates, or employers, for a given round. Withholdings mainly come from a matching algorithm not meeting the requirement of having dense and square preference datasets, which is most commonly the case in real-world examples. We want to measure withholdings because if a matching algorithm yields a high number of withholdings, they're be less match options for those candidates or employers that were withheld from the matching algorithm rounds.

Returning about to our first measurement, displacement, we need to consider how withholdings can affect the displacement calculation. In the past examples, we only calculated the average displacement for the candidates, or employers, who were given a match. We need to consider the cons of those candidates, or employers, who were withheld from receiving a match. To just ignore them in the average displacement calculation isn't enough. As briefly mentioned earlier in Chapter 3, those candidates, or employers, who don't receive a match for a specific round, will receive a less important match in a later round. Candidates, or employers, may not receive a match for a round because all of their preferences were already assigned to someone more fitting, which is where the *stability* of these matches comes into play. To stress how these withholdings cause candidates, or employers, to receive less important matches in later rounds, we need to add a penalty cost to the displacement value if they were withheld for a specific round. In other words, if a candidate or employer is withheld, the next round that they receive a match will have an additional cost to their displacement. This penalty cost is the number of preferences that candidate, or employer, holds.

## 4.2   Experimental Datasets

For our first experimental datasets, we simulated the function of a traditional matching market, we generated preferences for *n* candidates and *m* employers. Each candidate and employer are randomly assigned two attribute values drawn from the normal distribution with a mean of 0 and a standard deviation of 1. Each candidate and employer, moreover, has randomly assigned one of two types with even odds. Type-1 candidates and employers find the first attribute twice as important as the second. Type-2 candidates and employers find the second attribute twice as important as the first. Candidates rank 10 random employers by the resulting utility, and the employers, receiving their applications, rank the candidates in their pool by their utility. Having a utility function makes these preference datasets more realistic, rather than just being completely random. These preference datasets are represented as a matrix, where each row represents individual candidates, or employers; and the columns represent their decreasing preferences. Examples of these datasets can be found in the figures in Chapter 3. We repeated this dataset generation process for different amounts of candidates and employers, such as 10 vs 100, 50 vs 100, 100 vs 100, 110 vs 100, and 150 vs 100 (Number of candidates vs Number of employers). The number of employers is kept constant at 100 to ensure the *displacement* penalty is the same across the different datasets.

Using the experimental datasets, we ran all three algorithms and evaluated them using the measurements described in Chapter 4.1. For LMF-MMDAA and Mixed MMDAA, we set the max number of rounds to be the same as the ones used for the Normal MMDAA to converge. This is because the Normal MMDAA terminates at a much earlier number of iterations/rounds, and we want to compare each algorithm on a round-by-round basis. For example, the Normal MMDAA may converge at around 18 iterations/rounds, where each candidate has 10 preferences, while the LMF-MMDAA may converge at around 143 iterations/rounds, where each candidate has 100 preferences (through LMF). Running the LMF-MMDAA for this long isn't very practical, since most of the important matches are found within the first 10 or so matches/rounds. For this reason, we don't need to run the full 143 rounds on the LMF-MMDAA, 18 iterations/rounds will suffice. The below figures show the average displacement and withholdings for employers and candidates on different datasets over many iterations/rounds for various multi-match algorithms.

## 4.3  Employers' Average Displacement

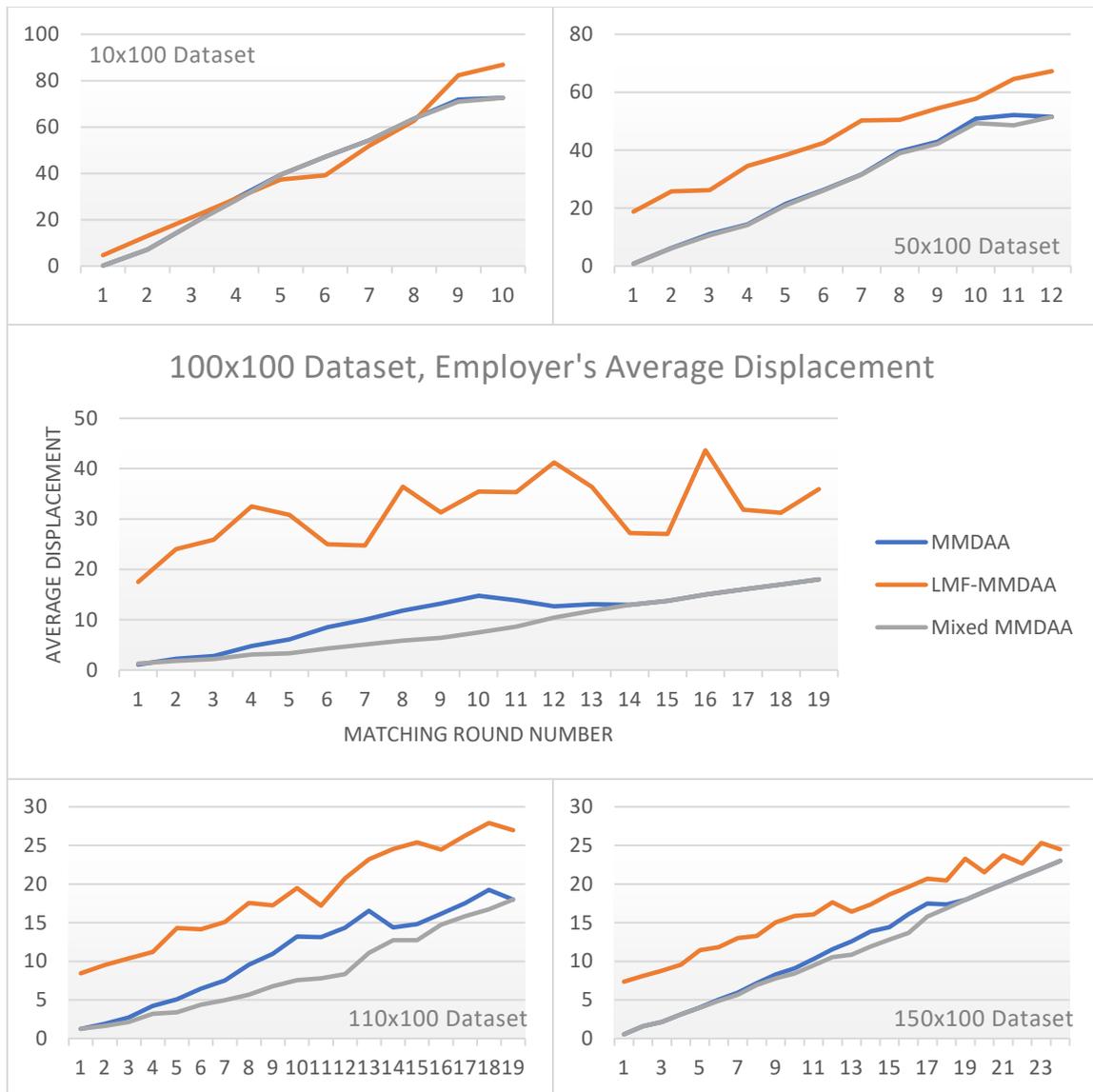

Figure 4.1: Employers' Average Displacement Results

    As expected, Normal MMDAA has an employers' displacement that is always in-between LMF-MMDAA and Mixed MMDAA, across all 5 of the datasets. This shows that the employers' displacement is consistently worse than Mixed MMDAA and better than LMF-MMDAA. Note that in the first Dataset (10x100), we see that all three algorithms have a very close employers' displacement. This is because there are only 10 candidates in that dataset, causing lots of withholdings on the employer's side, which creates displacement penalties as explained in the technical section. This lack of candidates is also why the employers' displacement values are much higher overall in the datasets that have less candidates than employers. In the datasets with excess candidates, the employers' displacement is lower because they have more candidates to choose from, some having higher preference rankings than others. Across all 5 datasets, the employers' displacement starts off with a low value, and gradually increases as we progress through matching rounds. This is because 1) later matches naturally have a lower match accuracy (since all the better matches with higher preference rankings are

found in earlier rounds), and 2) more withholdings occur during later matching rounds, which create displacement penalties.

Focusing on the LMF-MMDAA, I found that the employers' displacement is always higher compared to the other two algorithms, across all 5 of the datasets. This is because we are using LMF-generated preference datasets instead of the original preference datasets, which has a loss of preference ranking accuracy between the candidates and employers. Also, withholdings occur during later matching rounds, which create some displacement penalties. As explained earlier, LMF-MMDAA tends to yield high displacement, but the benefit of using this algorithm is that it achieves low withholding, which we will see later. In the 10x100 dataset, there are so few candidates that the LMF-generated preference datasets are very similar to the original preference datasets, due to the lack of data. This resulted in the LMF-MMDAA yielding a similar employers' displacement compared to the other algorithms. In the other datasets, LMF-MMDAA was able to generate more meaningful data, which can be shown by the change in the employers' displacement.

As for Mixed-MMDAA, the employer's displacement was the lowest out of the three different algorithms, across each of the 5 datasets. This is expected since the Mixed-MMDAA uses the match results from the original preference datasets, which has a high preference ranking accuracy between candidates and employers, and it uses the match results from the LMF-generated preference datasets *only* when there is an employer being withheld (as explained in Chapter 3.4). We have the low displacement from the Normal MMDAA algorithm, and when there are withholds, we use the LMF-MMDAA matches rather than assigning a displacement penalty for the withhold. Note that the Mixed-MMDAA is most effective when there are many candidates to choose from. As stated earlier, one of the main reasons why the employers' displacement gets high is because of withholdings, since they cause displacement penalties. If there is a low number of candidates, such as in the 10x100 and 50x100 datasets, then at most there will be 10 employers in the 10x100 dataset, and 50 employers in the 50x100 dataset, that will receive a match for a given round. The Mixed-MMDAA won't be able to fill matches for the employers being withheld by using the LMF-MMDAA match results, simply due to the lack of candidates. Therefore, the Mixed-MMDAA has a similar employers' displacement to the Normal MMDAA, because there are no available substitutions for withholdings. In the other datasets, these employer withholdings can be substituted through the Mixed MMDAA, since we have excess candidates.

### 4.4     Candidates' Average Displacement

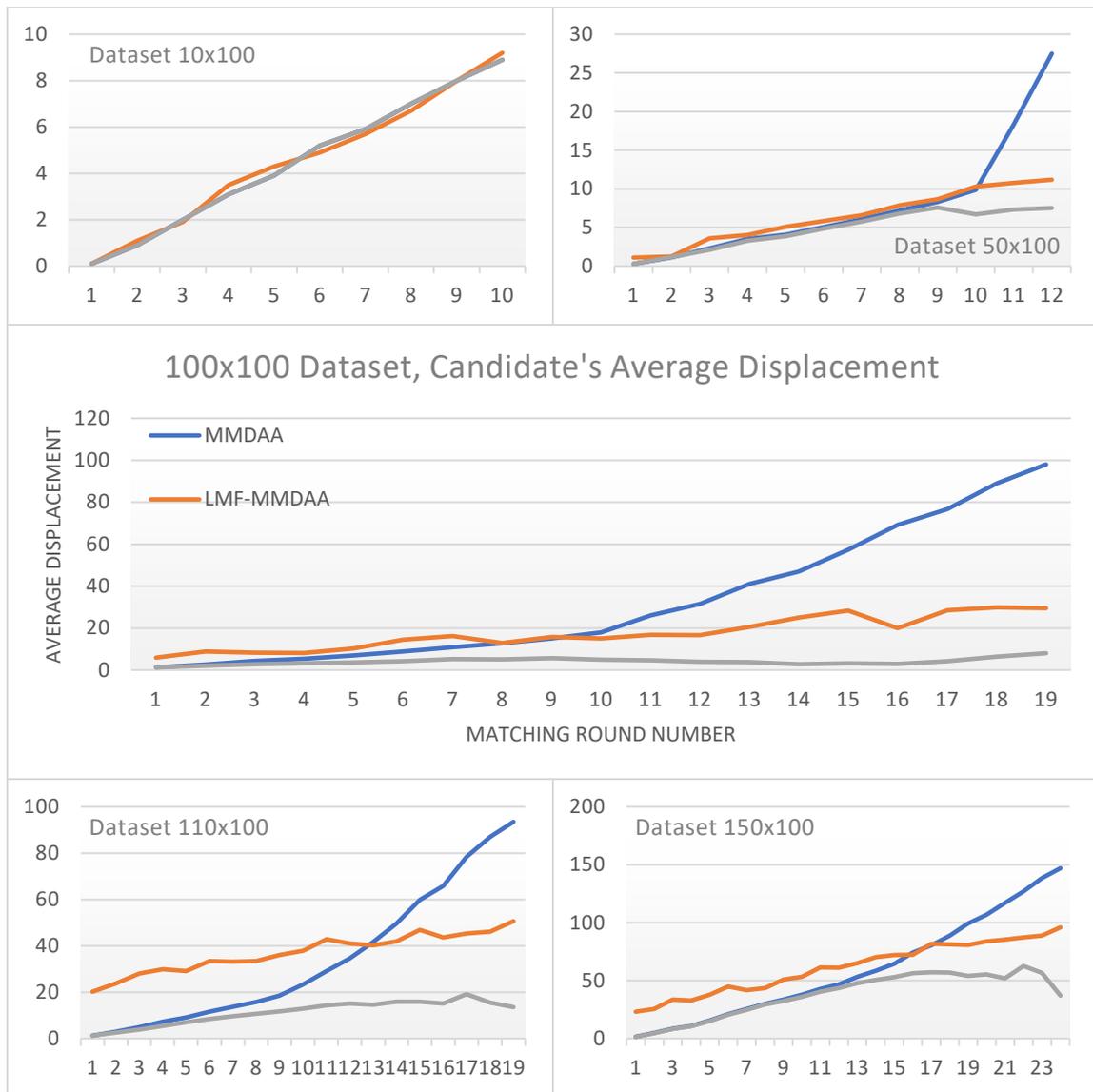

Figure 4.2: Candidates' Average Displacement Results

    Unlike Normal MMDAA's employers' displacement, Normal MMDAA has a candidates' displacement that isn't always in-between LMF-MMDAA and Mixed MMDAA. On every dataset, besides the 10x100 dataset, we can see that the candidates' displacement starts off in-between the two other algorithms, but eventually shoots up past LMF-MMDAA. This is because we have candidates being withheld, due to lack of preferences, which cause displacement penalties. Withholdings are more common during later matching rounds. The overall candidates' displacement value is lowest when we have a small number of candidates, because those candidates will have more employers to choose from, some having higher preference rankings than others. Note that in the first Dataset (10x100), we don't see the curve for the Normal MMDAA candidates' displacement. This is because it is completely underneath the Mixed-MMDAA. In other words, the Mixed-MMDAA and the Normal MMDAA have the same candidates' displacement. We can infer that since the candidates' displacements are equivalent, then the Mixed-MMDAA match results should be the same as the Normal MMDAA match

results. This is due to the Mixed-MMDAA not having to substitute a withholding with an LMF-MMDAA match, which means that the Mixed-MMDAA must have solely used the Normal MMDAA match results because there were no withholdings. Logically this makes sense, because there are only 10 candidates against 100 employers, so there is a high probability that no candidates will be withheld for a matching round. Across all 5 datasets, the candidates' displacement starts off with a low value, and gradually increases as we progress through matching rounds. This is because 1) later matches naturally have a lower match accuracy (since all the better matches with higher preference rankings are found in earlier rounds), and 2) more withholdings occur during later matching rounds, which create displacement penalties.

The LMF-MMDAA tends to have a relatively high candidates' displacement. This is because we are using LMF-generated preference datasets instead of the original preference datasets, which has a loss of preference ranking accuracy between the candidates and employers. The LMF-MMDAA's candidates' displacement is initially worse than the Normal MMDAA's candidates' displacement, but it steadily increases as compared to the Normal MMDAA's spiked increase in displacement. The candidates' displacement for the LMF-MMDAA is increasing without a sudden spike because of the low rate of withholdings which comes from using the LMF-generated preference datasets. In the 10x100 dataset, there are so few candidates that the LMF-generated preference datasets are very similar to the original preference datasets, due to the lack of data. This resulted in the LMF-MMDAA yielding a similar candidates' displacement compared to the other algorithms. In the other datasets, LMF-MMDAA was able to generate more meaningful data, which can be shown by the change in the candidates' displacement.

The candidates' displacement for the Mixed-MMDAA, as expected, was the lowest out of the three different algorithms, across each of the 5 datasets. We have the low displacement from the Normal MMDAA algorithm, and when there are withholds, we use the LMF-MMDAA matches rather than assigning a displacement penalty for the withhold. The Mixed-MMDAA is most effective when there are many candidates. The 10x100 and 50x100 datasets have a small number of candidates, which means there won't be many withholdings for the Mixed-MMDAA to fill in with LMF-MMDAA matches in order to improve the candidates' displacement. In later matching rounds some withholdings occur, which is why the Mixed-MMDAA remains to have a moderately low displacement as compared to the Normal MMDAA. This can be seen in the 10x100 and 50x100 datasets. In the other datasets, the Mixed-MMDAA continues to yield a low candidates' displacement.

## 4.5 Employers' Average Withholdings

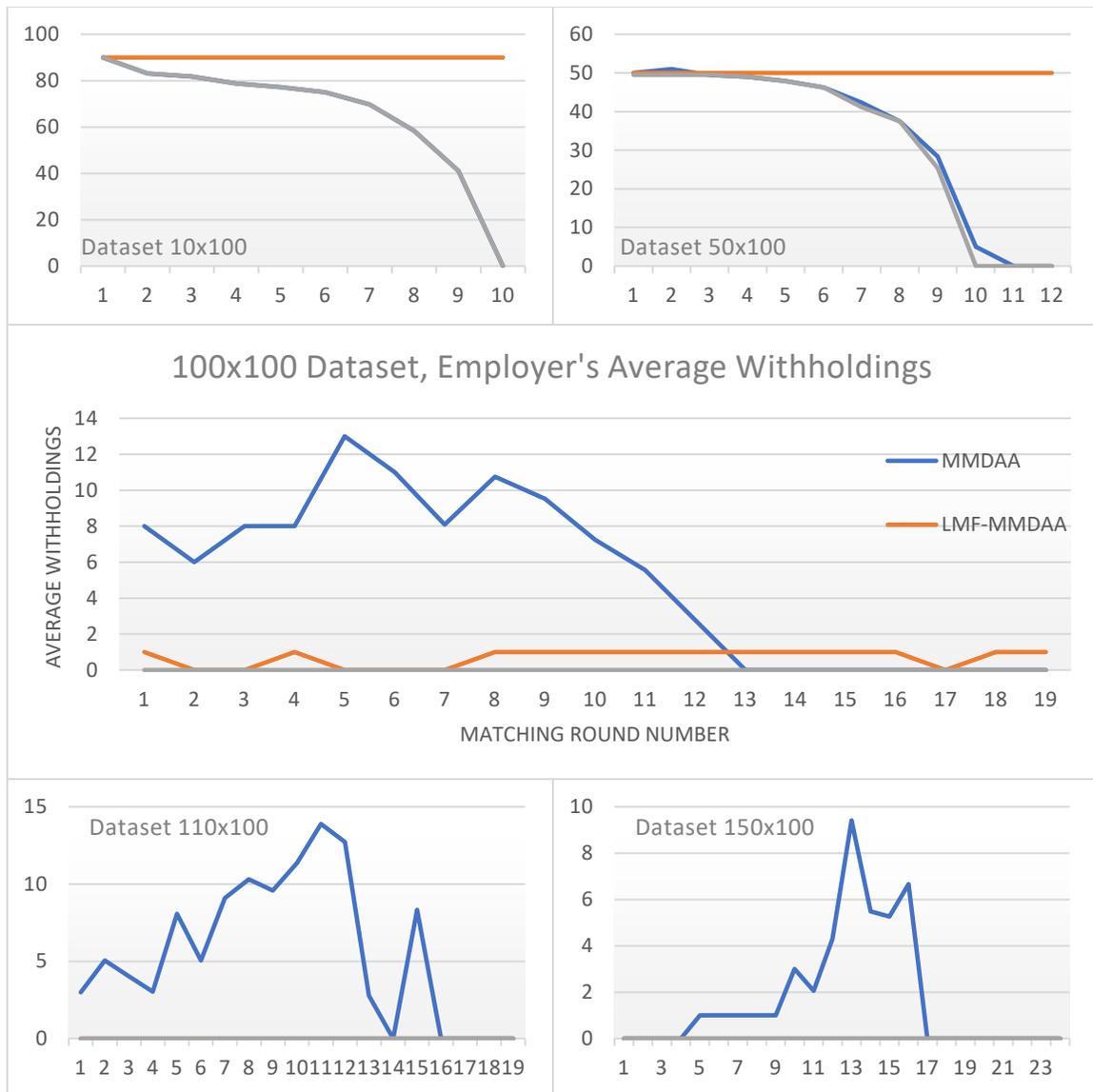

Figure 4.3 Employers' Average Withholdings Results

    For the Normal MMDAA, we see that for the 10x100 and 50x100 datasets, the employers' withholdings gradually decreases. Using the 10x100 dataset as an example, we would have 10 candidates and 100 employers for the first matching round. This would indicate that after the 10 candidates have been matched, 90 out of the 100 employers would be withheld on the first round. As we progress to later rounds, some of the employers would have matched with everyone on their preference list, which means they are taken out of the algorithm, lowering the average withholdings for employers. For any round in the 10x100 dataset, there will be at most 10 matched employers since there are only 10 available candidates. For example, on the first round we have 90/100 employers withheld (90%), and on the $5^{th}$ round we have around 34/44 employers withheld (77%). This pattern is also followed on the 50x100 dataset. As for the other three datasets, the 100x100, 110x100, and 150x100, the Normal MMDAA's employers' withholdings spike up for a bit, then start to decrease. This occurs due to the natural complexity of the matching algorithm. At any given round, all of an employer's list of preferred candidates may

have been matched with another employer, which forces a withhold. This is more likely to occur in later rounds, as the preferences are getting removed, which reduces the number of choices an employer has to match with a candidate. Also note that across all 5 datasets, the Normal MMDAA's employers' withholdings eventually reaches 0, because all preferences have been matched by the last round.

The LMF-MMDAA has some interesting results for the employers' withholdings. The LMF-MMDAA is supposed to lower the rate of withholdings, so why is it higher than the Normal MMDAA on the 10x100 and 50x100 datasets? The LMF-MMDAA did, in fact, have a lower *rate* of withholdings. These figures do not correspond to the number of withholdings per round; it corresponds to the *average* number of withholdings per round. In other words, it takes the number of withholdings, and divides it by the total number of employers that remain in the algorithm. For the Normal MMDAA, we see the employers' withholdings gradually decreasing because the number of employers that remain in the algorithm decreases as they match with everyone on their preference list. With LMF-MMDAA, this is not the case. When we run the LMF-generated preference datasets through the matching algorithm, the employers won't be removed from the algorithm at the same time (matching round) as Normal MMDAA. This is because the preference datasets are denser and contain much more preferences per employer. For an employer to be removed from the algorithm, it would have to match with everyone on this new and longer preference list, which occurs at a much later round. Therefore, the LMF-MMDAA has a constant average employers' withholdings for the 10x100 and 50x100 datasets, because all 100 employers are still in the algorithm for the listed number of matching rounds. As for the 100x100 dataset, LMF-MMDAA has it's occasional few withholdings due to the algorithm removing preferences as it progresses through the rounds, which lowers an employer's total number of choices to match with a candidate. (Having a small number of choices can cause a withhold when other employers have matched with all of your candidates). The 110x100 and 150x100 datasets have a 0 employers' withholding because there is an excess of candidates to match with, and the LMF-generated preference datasets ensures that there is a preference for every candidate.

Mixed MMDAA's employers' withholdings follows a similar pattern to both, the Normal MMDAA, and the LMF-MMDAA. For the datasets with a small number of candidates, 10x100 and 50x100, the Mixed MMDAA employers' withholdings follows the same trajectory has the Normal MMDAA. This is because the Mixed MMDAA always uses the Normal MMDAA's matches as much as possible and fills in any withholdings with LMF-MMDAA matches. For these datasets with a small number of candidates, there's nothing the Mixed MMDAA can do to fill in the withholdings because of the lack of candidates, which means the result of the Mixed MMDAA is basically the same as the Normal MMDAA. For the remaining 3 datasets, 100x100, 110x100, and 150x100, there is enough candidates for each employer, so any withholdings that the Normal MMDAA ran into, they can be filled by the Mixed MMDAA. This resulted in the Mixed MMDAA having a 0 employers' withholdings.

## 4.6   Candidates' Average Withholdings

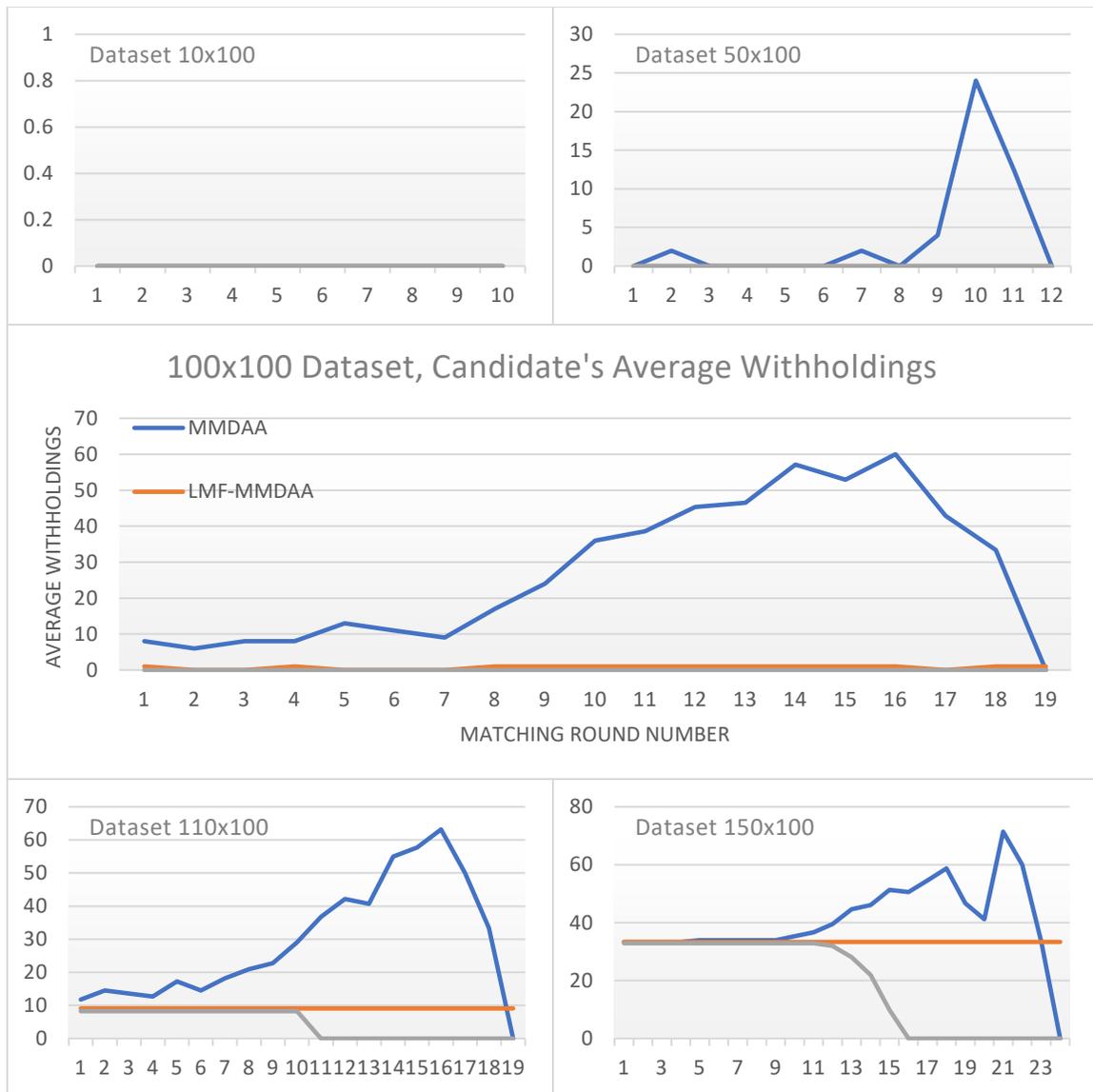

Figure 4.4: Average Candidates' Withholdings Results

    The Normal MMDAA for the candidates' withholding is a bit different compared to the employers' withholding. On the 10x100 dataset, we have a 0 candidates' withholding. This is simply because we only have 10 candidates against 100 employers, and it's not likely for one of those 10 candidates to be withheld (although it is possible). On the 50x100 dataset, we have a low average candidates' withholdings due to the nature of the matching algorithm. Each round the algorithm removes preferences, so by the 10th round, each candidate may have just one preference remaining (since each candidate starts with 10 preferences). In this scenario, many candidates may only have one preference, and there is no guarantee that this preference is unique to every candidate. In other words, two candidates can share this single preference, forcing one of them to be withheld. This problem is also the reason behind the high candidates' withholding for the 100x100, 110x100, and 150x100 datasets. More specifically on the 110x100 and 150x100 datasets, we have an excess number of candidates, which also raises our candidates' withholding

value. Note that across all 5 datasets, the Normal MMDAA's candidates' withholdings eventually reaches 0, because all preferences have been matched by the last round.

The LMF-MMDAA had some similar behavior to the Normal MMDAA in terms of its candidates' withholdings for the first two datasets, the 10x100 and 50x100. In the 10x100, the results are the same, LMF-MMDAA also achieved a 0 candidates' withholdings since there are only 10 candidates. For the 50x100 datasets, the low candidates' withholdings that came from the Normal MMDAA were completely patched up by the LMF-MMDAA, yielding a 0 candidates' withholding, due to the dense LMF-generated preference datasets. As for the 100x100 dataset, LMF-MMDAA has it's occasional few withholdings due to the algorithm removing preferences as it progresses through the rounds, which lowers a candidate's total number of choices to match with an employer. (Having a small number of choices can cause a withhold when other candidates have matched with all of your employers). For the 110x100 and 150x100 datasets, we see that the LMF-MMDAA's candidates' withholdings are constant. This is caused by the same reason for the LMF-MMDAA's employers' withholdings being constant on the 10x100 and 50x100 datasets – the LMF-generated preference datasets make it harder for candidates to be removed from the algorithm at earlier matching rounds. Since candidates aren't being removed from the algorithm, and there's always a set number of candidates forced to be withheld due to lack of employers, there will be a constant average number of candidates being withheld per round. For the 110x100 dataset, 10 candidates must to be withheld each round, 10/110 (or 9%), and for the 150x100 dataset, 50 candidates must be withheld each round is 50/150 (33%).

Mixed MMDAA's candidates' withholdings follows a similar pattern to the LMF-MMDAA. For the 10x100, 50x100, and 100x100 datasets, the Mixed MMDAA candidates' withholdings follows a similar trajectory as the LMF-MMDAA. This is because the Mixed MMDAA always uses the Normal MMDAA's matches as much as possible and fills in any withholdings with LMF-MMDAA matches. In the 10x100 dataset, there are no withholdings for Mixed MMDAA to fill in, but in the 50x100 dataset, Normal MMDAA had a few withholdings which were able to be filled in by the Mixed MMDAA using the LMF-MMDAA matches, which resulted in a 0 candidates' withholding. For the 100x100 dataset, there were many withholdings from the Normal MMDAA. However, the Mixed MMDAA was able to use the LMF-MMDAA to fill in all of them, for each round, yielding a 0 candidates' withholding. The 110x100 and 150x100 datasets were the most interesting. The Mixed MMDAA couldn't achieve a 0 candidates' withholding at the start simply because there are just not enough employers for each candidate, forcing candidates to be withheld. However, eventually (in later matching rounds) some of the candidates have matched with every one of their preferences, which means they get pulled out of the algorithm. This implies that the total number of candidates will drop to 100, or lower, during later matching rounds, which means there is enough employers for every candidate, giving the Mixed MMDAA the ability to fill in all withholds with LMF-MMDAA matches. Because of this, Mixed MMDAA is able to reach a 0 candidates' withholdings.

## 4.7    LMF-MMDAA Compared to Original Preferences

The above figures/experiments compared the LMF-MMDAA to the input preferences, as in the preferences that resulted from Low-Rank Matrix Factorization. As mentioned earlier, the original preferences are overwritten in the LMF-MMDAA, and there is no guarantee that they are considered first when running the MMDAA. Below are a few figures regarding how the LMF-MMDAA results compare to the original preference rankings.

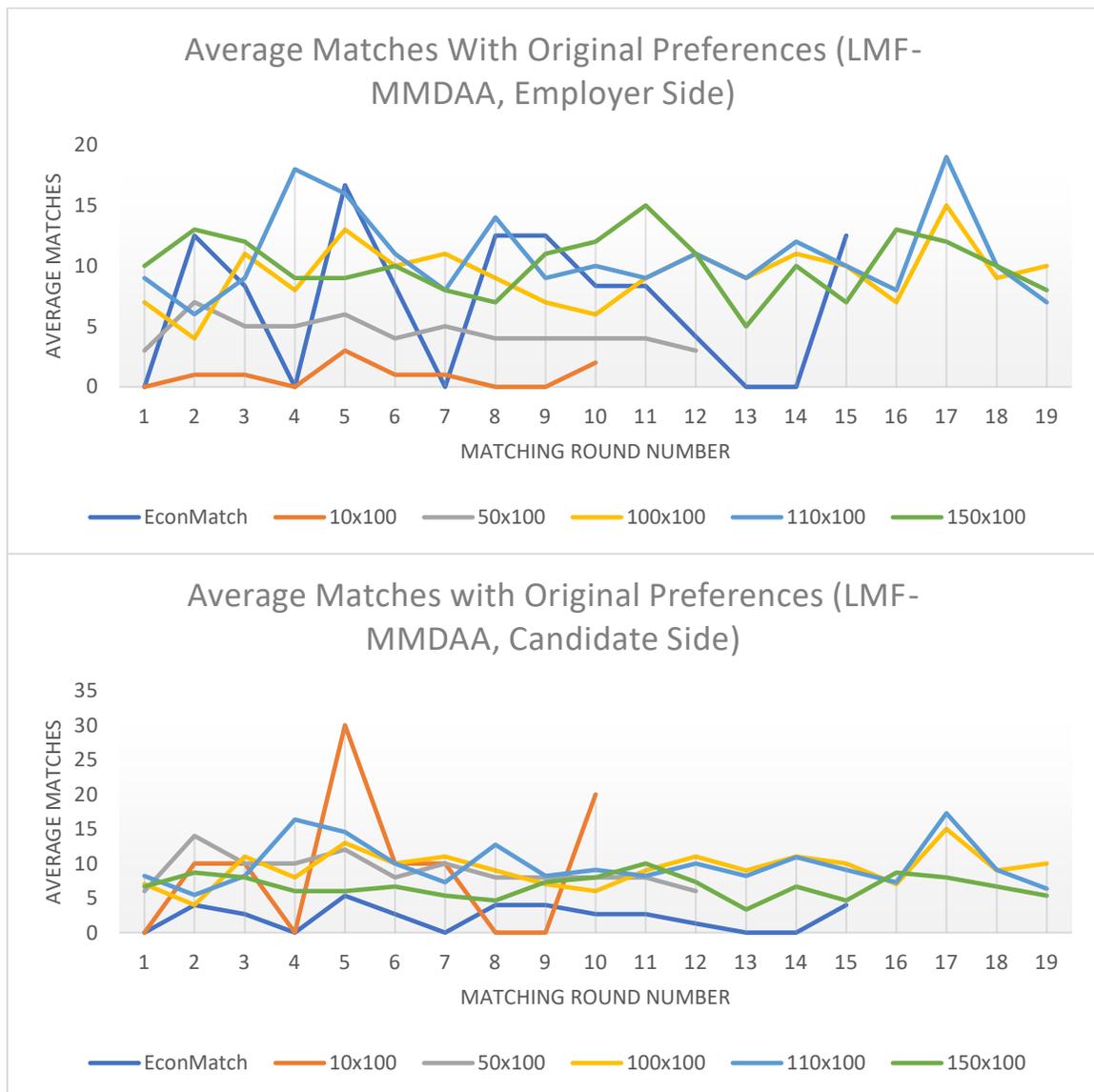

Figure 4.5: Average Matches against Original Preferences (LMF-MMDAA)

For each of the synthetic datasets, we recorded the average number of LMF-MMDAA matches containing an original preference, for both candidates and employers (before applying LMF). Figure 4.5 shows that across all of the datasets, roughly no more than 20% of candidates and employers receive a match that contained an original preference, for each matching round. 80% of the candidates and employers receive matches that have a LMF-generated preference, which isn't ideal in the sense that users should match with who they chose originally. The same number of rounds that were used in the average displacement and average withholdings figures were used in Figure 4.5, capped at 19. EconMatch is a dataset composed of real data that will be explained and analyzed in more detail in Chapter 6.

# Chapter 5

# Simulating a Job Market to Evaluate Vacancy Improvements

## 5.1 Job Market Simulation Algorithm

To ensure that our algorithms are making the job market more efficient, we created a job market simulator. One of the main reasons the MMDAA was created is to reduce the number of unhired candidates and unfilled job positions. The MMDAA is used to create stable one-to-one matches for employers to find highly potential, recommended candidates for their position. Having a recommended candidate list for an open position reduces the amount of resources an employer has to spend looking for a new employee, such as time and money. Employers wouldn't want to waste resources on going after/interviewing candidates that wouldn't be a good fit in the first place (because candidates may have interests in other employers, which is what the MMDAA will utilize to find optimal matches). Our job market simulator shows the results of what would happen if 1) employers offered positions to candidates based off of their direct preferences, without the use of the MMDAA, 2) employers offered positions to candidates based off of their results of any of the three MMDAA's. We want to ensure that the number of jobless candidates and unfilled positions is minimal when using an MMDAA, compared to having employers chase candidates by themselves, which helps create a better job market for the economy.

| Employer ID | Pref 1 | Pref 2 | Pref 3 | Pref 4 | Pref 5 | Pref 6 | Pref 7 | Pref 8 | Pref 9 | Pref 10 | Pref 11 | Pref 12 | Pref 13 | Pref 14 |
|---|---|---|---|---|---|---|---|---|---|---|---|---|---|---|
| Employer 1 | 2 | 7 | 13 | 11 | 14 | 1 | 3 | 6 | 5 | 8 | 4 | 9 | | |
| Employer 2 | 3 | 12 | 1 | 2 | 5 | 7 | 13 | | | | | | | |
| Employer 3 | 12 | 8 | 2 | 6 | 1 | 3 | 11 | | | | | | | |
| Employer 4 | 14 | 10 | 13 | 2 | 8 | 9 | 12 | 6 | 3 | 1 | 5 | | | |
| Employer 5 | 6 | 11 | 5 | 14 | 7 | 10 | 4 | 9 | 1 | 2 | 3 | 8 | | |
| Employer 6 | 9 | 2 | 13 | 5 | 10 | 4 | 14 | 6 | 12 | 1 | 11 | 3 | | |
| Employer 7 | 13 | 11 | 12 | 8 | 1 | 6 | 4 | 10 | 7 | 2 | 5 | 3 | 9 | 14 |
| Employer 8 | 1 | 5 | 12 | 2 | 4 | 8 | 14 | 13 | 3 | 10 | 11 | 6 | 7 | 9 |
| Employer 9 | 14 | 7 | 8 | 11 | 2 | 4 | 5 | 3 | 6 | 1 | 10 | 9 | 12 | |
| Employer 10 | 11 | 3 | 13 | 2 | 4 | 1 | 7 | 5 | 14 | 10 | 12 | 6 | | |

Figure 5.1: Simulated Job Market without MMDAA

To make our job market simulator realistic, we designed it with the help of an Economics professor, Andrew Johnston. The job market simulator is as follows: We have 3 classes of employers, 1) high-class employers, 2) medium-class employers, 3) low-class employers. For our testing datasets, we partitioned every third of the employers to each of the three classes. In other words, the first 33 employers are high-class, the next 33 employers are medium-class, and the last 34 employers are low-class. In the above figure, we are showing an example of how we partition a dataset for the job market simulator with 10 employers. Each third of the employers, who each have their own Employer ID, are categorized into high-class (green), medium-class (yellow), and low-class (red). Once the employers are placed in their respective classes, we simulate how they offer positions to candidates based off of the employers' class type.

There will be 3 position offering sessions. In other words, each employer will have 3 rounds/chances to offer a position to candidate. High-class employers will offer positions to candidates that are most preferred. Medium-class employers will offer positions to candidates that are preferred, but not most preferred. More specifically, we have the medium-class employers offer positions to candidates who are after the top 33% of preferred candidates. Using the above figure as an example, we see that employer 4 is going to offer a position to its 4$^{th}$ preferred candidate first, which is candidate #2. For the remaining two rounds, employer 4 will linearly offer positions to whoever is next in line, which in this case would be the 5$^{th}$ and 6$^{th}$ preferred candidates, candidate #8 and candidate #9. Medium-class employers offers positions to candidates in this specific fashion because it simulates a more realistic job market. Medium-class employers know that their highly qualified, more preferred candidates will probably receive better offers by an employer with a higher class. For this reason, we have low-class employers offering positions to candidates who are after the top 66% of preferred candidates. The above figure color-coordinates which candidates each employer will offer positions to, based off of the employers' class.

On the candidate side, the simulator will have all candidates accept the first offer they receive (if they even receive an offer). A candidate cannot accept more than one offer, he/she must decline any extra offers. For example, in the above figure, we have employer 1 offering a position to candidate #2 during the first out of the three rounds. Candidate #2 will accept this offered position, and must decline any offer it receives in the current, or later rounds. In the same round, employer 4 and employer 7 will offer a position to candidate #2, which he/she will have to decline. In the third and last round, candidate #2 also declines the offer from employer 3.

Our job market simulator measures a new metric, *vacancy*. For employers, *vacancy* measures the number of unfilled positions, after a specific amount of job offering rounds. For candidates, *vacancy* measures the number of jobless candidates, after a specific amount of job offering rounds. Vacancy will always be at its maximum after the first job offering round, and vacancy will lower as we continue to perform more job offering rounds, because 1) employers receive more chances to fill a position by offering jobs to candidates, and 2) candidates receive more chances to accept an offer.

In the above figure, the vacancy for employers after the first job offering round would be 30%. This is because 3 out of the 10 employers didn't fill their position. Employers 4 and 7 both had their position declined by candidate #2 because candidate #2 has already accepted an offer from employer 1. Employer 8 didn't fill their position because candidate #10 accepted an offer from employer 6 first. During the second round of job offerings, employers 4, 7, and 8, will offer jobs to the candidate next in line in their preferences. All three of these employers will have their candidates accept the position, as those candidates have not received any other position in the current, or previous round. The vacancy for employers after the second job offering round would be 0%, because all employers have successfully filled their positions. No employer will participate in the third matching round, because they all have found their candidates. We can also calculate the vacancy for candidates by counting the number of jobless candidates after each matching round. Since there are 14 candidates and 10 employers in this example, we know that there will be at least 4 jobless candidates, simply due to the lack of employers.

The previous example was simulating a job market based off of employers' preferences on candidates. Now we want to simulate what would happen, in terms of *vacancy*, if we have the employers offer jobs to candidates based off of how they were matched from an MMDAA.

| Employer ID | Match 1 | Match 2 | Match 3 | Match 4 | Match 5 | Match 6 | Match 7 | Match 8 | Match 9 | Match 10 | Match 11 | Match 12 | Match 13 | Match 14 |
|---|---|---|---|---|---|---|---|---|---|---|---|---|---|---|
| Employer 1 | 20 | 8 | 4 | 26 | 1 | -1 | 53 | 94 | 93 | 60 | 10 | 64 | | |
| Employer 2 | 68 | 12 | 40 | -1 | -1 | -1 | 92 | | | | | | | |
| Employer 3 | 94 | 28 | 58 | 24 | 78 | 36 | 75 | | | | | | | |
| Employer 4 | 79 | 43 | 16 | 75 | -1 | 40 | 88 | 45 | 64 | 95 | 2 | | | |
| Employer 5 | 51 | 3 | 74 | 36 | 93 | 5 | 25 | 53 | 75 | 32 | -1 | 17 | | |
| Employer 6 | 18 | 27 | 32 | 69 | 52 | 21 | 46 | 42 | 50 | -1 | -1 | 11 | | |
| Employer 7 | 96 | 49 | 73 | 98 | 100 | 1 | 4 | 48 | 85 | 29 | 90 | 19 | 84 | 10 |
| Employer 8 | 35 | 25 | 15 | 14 | 40 | 72 | 41 | 38 | 10 | 57 | 77 | 60 | 76 | 64 |
| Employer 9 | 48 | 77 | 24 | 95 | 58 | 35 | 80 | 19 | 41 | 55 | 22 | 85 | 34 | |
| Employer 10 | 84 | 63 | 38 | 37 | 36 | 12 | 89 | 87 | 61 | 72 | 26 | 16 | | |
| Employer 11 | 64 | 11 | 63 | 2 | 65 | 34 | 100 | 74 | 81 | 73 | 79 | | | |
| Employer 12 | 77 | 58 | -1 | 7 | 3 | 41 | 50 | 39 | | | | | | |
| Employer 13 | 91 | 99 | 25 | 32 | 75 | 16 | 24 | 49 | 56 | 11 | 81 | 10 | 77 | |
| Employer 14 | 62 | 68 | 45 | 97 | 24 | 100 | 56 | 15 | 72 | 18 | 99 | 25 | | |
| Employer 15 | 32 | 33 | 56 | 57 | 83 | 4 | 3 | 58 | | | | | | |
| Employer 16 | 54 | 1 | 30 | 31 | 2 | -1 | 72 | 46 | | | | | | |
| Employer 17 | 36 | 22 | 95 | 50 | 81 | 46 | 62 | 91 | 6 | 47 | 73 | 33 | | |
| Employer 18 | 93 | 26 | 91 | 21 | 97 | 69 | 9 | 47 | 98 | -1 | 43 | | | |
| Employer 19 | 1 | 24 | 9 | 8 | 86 | 50 | 10 | 31 | 76 | 7 | 36 | | | |
| Employer 20 | 31 | 85 | 66 | 51 | 59 | 17 | 52 | 33 | 78 | | | | | |

Figure 5.2: Simulated Job Market with MMDAA

The above figure shows how the employers would offer positions to candidates if they first used an MMDAA to find stable on-to-one matches. Each employer would simply offer positions to the first three candidate match results they receive from an MMDAA. We don't need to consider the type of class an employer is, because an MMDAA considers preferences from both sides (candidates and employers), when finding the stable matches. Essentially, an MMDAA passively considers the type of class an employer is, because employer class will show based on the candidate's preference of that employer (candidates would tend to rank high-class employers as highly preferred, compared to middle-class or low-class employers). Because of this, the stable one-to-one matches that arrives as output from an MMDAA can be directly used by the employers to offer positions to candidates. In other words, we have employers always offering positions to candidates based off of their best matches (the top 3 matches). Note that in the above figure, employer 12 did not receive a match for the 3$^{rd}$ round of the MMDAA, so it will use its 4$^{th}$ match as the candidate to offer a job to during the 3$^{rd}$ job offering round. Also note that candidate #7, who couldn't match with employer 12, isn't found in any of the other 19 employers' match results because this is data from a 100x100 MMDAA. There are 80 more employers with match results, one of which holds the match for candidate #7, which is why employer 12 didn't receive a match for the 3$^{rd}$ round of the MMDAA.

Using the same experimental datasets, we ran all three MMDAA's to obtain all of the match outputs. We then evaluated them by using the job market simulator and vacancy measurement described above. In the below figures, "Real World" refers to the simulation where employers offer positions to candidates based off of their direct preferences, without the use of any MMDAA, as showed in the first example in Figure 5.1. The below figures show the average vacancy for employers and candidates on the different datasets over each of the three job offering rounds, for various multi-match algorithms.

## 5.2    Employers' Average Vacancy

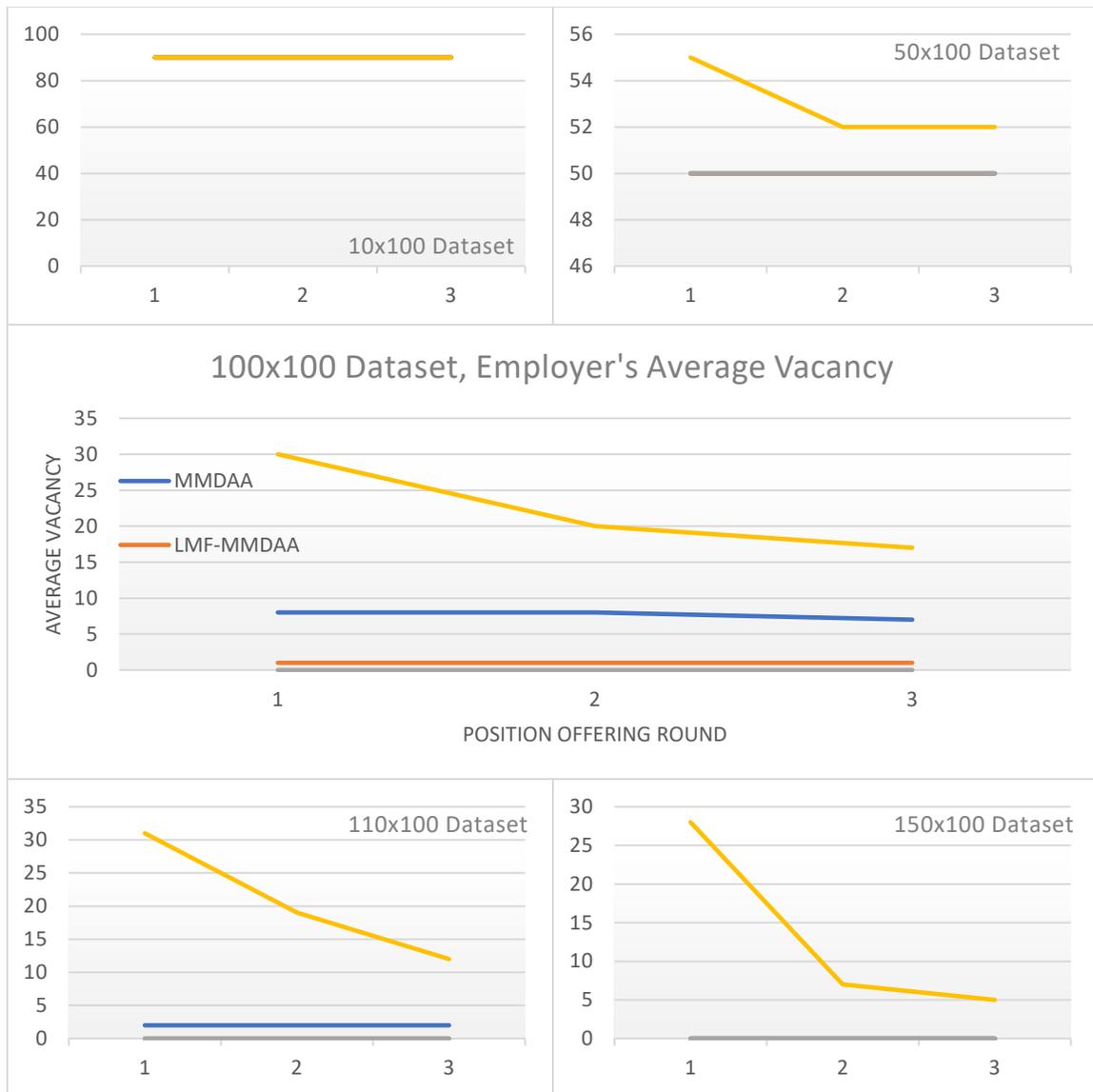

Figure 5.3: Employers' Average Vacancy Results

As expected, we have the Real World simulation yielding the highest employers' average vacancy across all five of our testing datasets. "Real World simulation" refers to employers offering positions to candidates based off of their direct preferences, without the use of any MMDAA, as showed in the first example of Figure 5.1. The main reason why the Real World simulation has the highest average vacancy is because the employers are offering positions to candidates based off of their preferences, which means that this is not a one-to-one job offering. In other words, many employers can be offering a position to the same candidate, which forces the candidate to decline the extra positions, creating vacancy for the employers.

Note that in the 10x100 dataset, all three MMDAA's simulation and the Real World simulation have the same average vacancy simply due to lack of candidates, because there will always be at most 10 employers who will find a candidate to accept a position, while the other 90 employers will be a part of the vacancy. The same consequence of lack of candidates can also be found in the 50x100 dataset, however, the Real World simulation is worse compared to any of the three MMDAA's simulation. As we know, the Normal MMDAA guarantees stable one-to-one matches, which helps reduce the amount of vacancy. The problem with the Normal MMDAA is that while it guarantees stable one-to-one matches, it doesn't guarantee that everyone gets a match, due to withholdings. For this reason, we see that across all five datasets, a simulated job market where employers use the results from a Normal MMDAA for offering positions to candidates has a lower average vacancy compared to having the employers offer positions based off of their bare preferences.

   The 100x100 dataset has the most interesting results: 1) the Real World simulation has the highest average vacancy due to employers offering positions to the same candidate, 2) the Normal MMDAA simulation has an moderately high average vacancy due to a high rate of withholdings in the algorithm itself, 3) the LMF-MMDAA simulation has a low, but not 0, average vacancy due to a few withholdings in the algorithm itself, and 4) the Mixed MMDAA simulation has a 0 average vacancy due to no withholdings in the algorithm itself. Note that even though the LMF-MMDAA simulation and the Mixed MMDAA simulation have a similar/same average vacancy in the 100x100, 110x100, and 150x100 datasets, the Mixed MMDAA has a more accurate set of matches for employers to follow (the displacement metric). In the 110x100 and 150x100 datasets, all three MMDAA's simulation have roughly the same vacancy due to an excess of candidates. Most/all of the 100 employers shouldn't have a problem with finding a candidate to accept their position after 3 position offering rounds. This isn't the case for the Real World simulation, as employers are still potentially offering positions to the same candidate.

## 5.3 Candidates' Average Vacancy

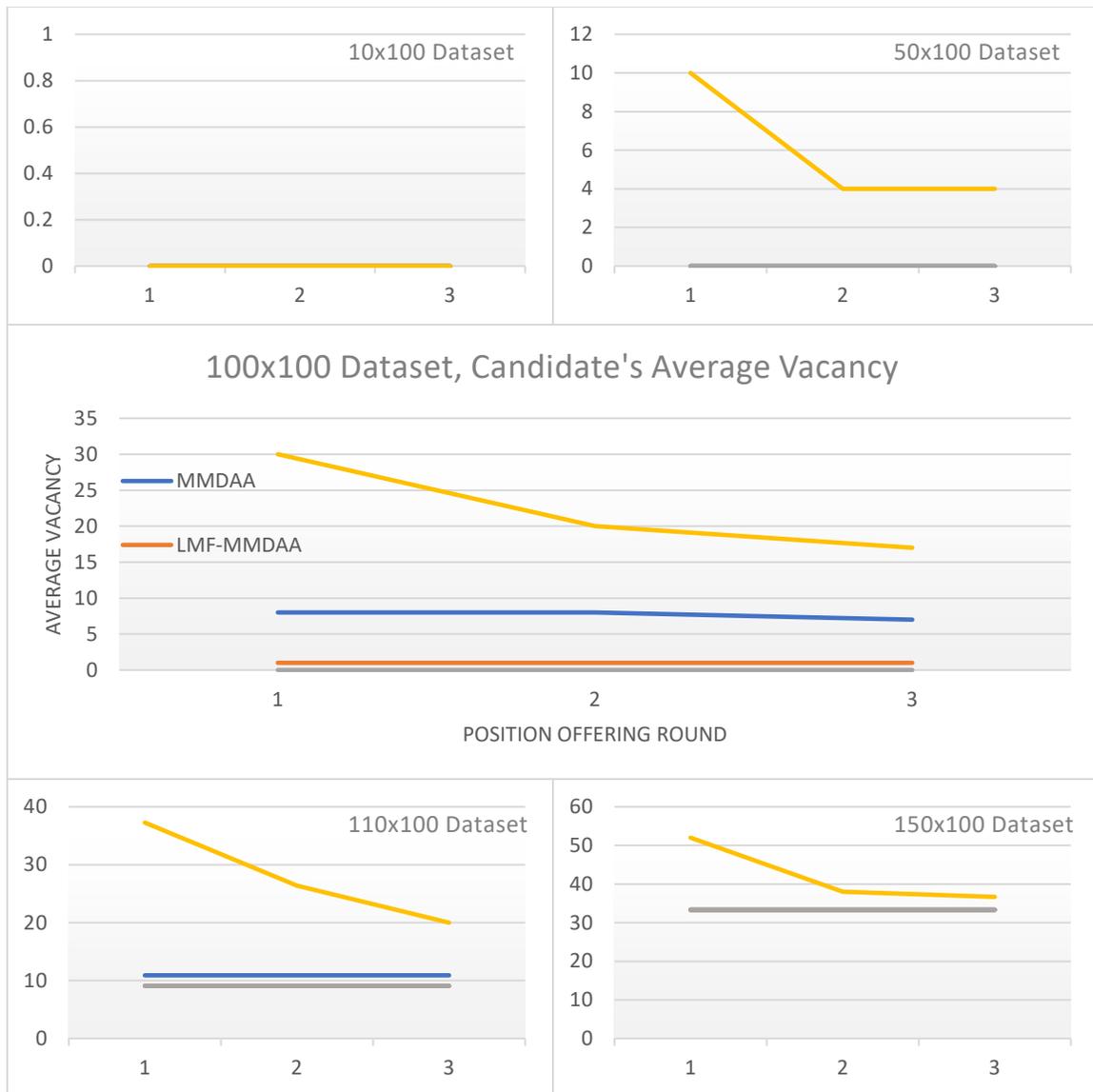

Figure 5.4: Candidates' Average Vacancy Results

Similar to the employers' average vacancy, the candidates' average vacancy was the highest in the Real World simulation, across all five of our testing datasets. In the Real World simulation, employers offer positions to candidates based off of their preferences and class, which means many candidates may receive multiple job offers while other candidates may receive no job offers. In other words, many employers can be offering a position to the same candidate, which forces some candidates to not receive a single job offer, creating vacancy for the candidates.

The average vacancy in the 10x100 dataset is 0 for all three MMDAA's simulations and the Real World simulation, simply due to an excess of employers. The Real World simulation was able to reach a 0 average vacancy due to all 10 candidates having a high chance of receiving at least one job offer, since there were 100 employers. All three MMDAA's simulations were able to achieve a 0 average vacancy in the 10x100 and 50x100 datasets due to excess employers causing a low average withholdings in the algorithms themselves. Note that the Real World simulator had a high average vacancy compared to any of the MMDAA's in the 50x100 dataset.

The 100x100 dataset has the most interesting results, similar to the employers' average vacancy. 1) the Real World simulation has the highest average vacancy due to employers offering positions to the same candidate, which leaves many candidates left without a job. 2) The Normal MMDAA simulation has a moderately high average vacancy due to a high rate of withholdings in the algorithm itself. 3) The LMF-MMDAA simulation has a low, but not 0, average vacancy due to a few withholdings in the algorithm itself, and 4) the Mixed MMDAA simulation has a 0 average vacancy due to no withholdings in the algorithm itself. In the 110x100 and 150x100 datasets, all three MMDAA's simulation have roughly the same vacancy due to a lack of employers. Even if every employer offers a job to a unique candidate (which is quite likely because MMDAA matches are one-to-one), there will be at most 100 candidates having a job offer, since there are only 100 employers, forcing candidate vacancy. This isn't the case for the Real World simulation, as employers are still potentially offering positions to the same candidate. Note that in the 150x100 dataset, the average vacancy is 33% because there are 50 candidates that didn't receive a job offer, out of the total 150 candidates in the simulator.

# Chapter 6

# Evaluating MMDAA's and Job Market Simulation on a Real Dataset

## 6.1 The EconMatch Dataset

Back in late 2018, we launched a website called "EconMatch." EconMatch is a website that allows users to make candidate or employer profiles. Candidate profiles can search for jobs that were posted on the AEAjobs website (a popular website for academic jobs in economics) and rank them in the order of preference. Employer profiles can search for jobs that they are responsible for, and view/rank candidates who have applied for that job. This website essentially allows us to gather preference data from candidates and employers, so we can run the Normal MMDAA on it and inform employers with the results, which would be recommended candidates for a position. We want to ensure that the Mixed MMDAA still produces the best results, even on real data, not just on experimental data. For this dataset, we managed to get 24 employers and 75 candidates.

## 6.2 MMDAA's on the EconMatch Dataset

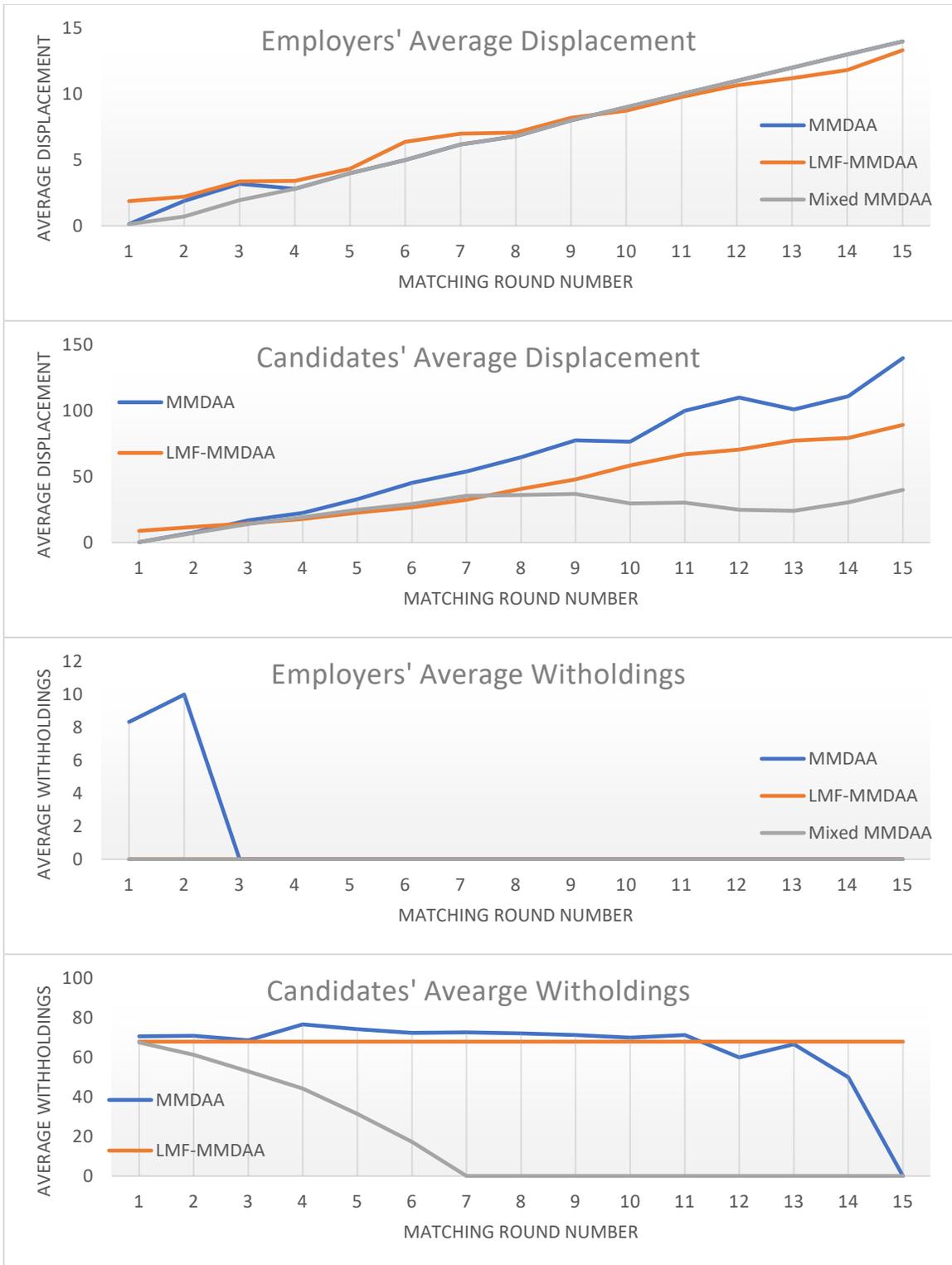

Figure 6.1: EconMatch Dataset MMDAA Results

On top of the high number of candidates, most of the preferences don't collide with each other. In other words, employers prefer very specific candidates that other employers don't have in common. Because of this, the preference datasets are already mostly one-to-one.

The Normal MMDAA, LMF-MMDAA, and Mixed MMDAA, all have a similar employers' average displacement. This is because there aren't many employers compared to the number of candidates, which means employers are likely to not be withheld. We can verify that this is true by looking at the employers' average withholdings figure. If employers don't get withheld, then there will be no displacement penalty, which is why all 3 MMDAA's have a similar displacement.

For the employers' average withholdings, the Normal MMDAA is the only algorithm that doesn't have 0 average withholdings for the earlier, more important, matching rounds. This is simply due to employers not having a wide variety of preferences. In other words, all of an employer's list of preferred candidates may have been matched with another employer, which forces a withhold for a matching round. LMF-MMDAA has 0 withholdings for the earlier rounds because it generates more preferences for the employers, so they won't force each other to be withheld. Mixed MMDAA uses these new LMF-MMDAA matches as fillings for all the withholdings in the Normal MMDAA matches, which also results in 0 withholdings for the earlier rounds.

The candidates' average displacement is more interesting compared to the employers' average displacement. This is because we have many candidates being withheld, which causes displacement penalties. The Normal MMDAA has the most withholdings, which results in it having the highest average displacement compared to the other two algorithms. The Mixed MMDAA has the least amount of withholdings, eventually diminishing to 0, which results in it having the lowest displacement. LMF-MMDAA has an average displacement that is in between the Normal MMDAA and Mixed MMDAA because it has a constant average withholdings that's in between the two other algorithms.

The Normal MMDAA initially has the highest average withholdings because of 1) a surplus of candidates (only 24 candidates will have a chance of being matched with the 24 employers, rest are withheld), and 2) candidates not having a wide variety of preferences. Candidates not having a wide variety of preferences raises withholdings because it's more likely for a candidate to have everyone on its preference list be matched with other candidates. LMF-MMDAA lowers the average withholdings to a constant value because the issue of candidates not having a wide variety of preferences is solved, which means the average withholdings is simply the result of the surplus of candidates. Note that the average withholdings for LMF-MMDAA is 68%, because only 24 candidates can match with the 24 employers, which means 51 candidates out of the 75 candidates will be withheld each round. The Mixed MMDAA also has this 68% average withholdings initially, but it lowers as the matching round progresses. This is because once a candidate has matched with everyone on it's preference list, it is taken out of the matching algorithm. This reduces the total amount of candidates in the algorithm, which is why the average withholdings lowers. Note that there is still is only 24 candidates matching with employers each round, and that the average withholdings only lowers due to the total number of candidates decreasing.

## 6.3 Job Market Simulation with the EconMatch Dataset

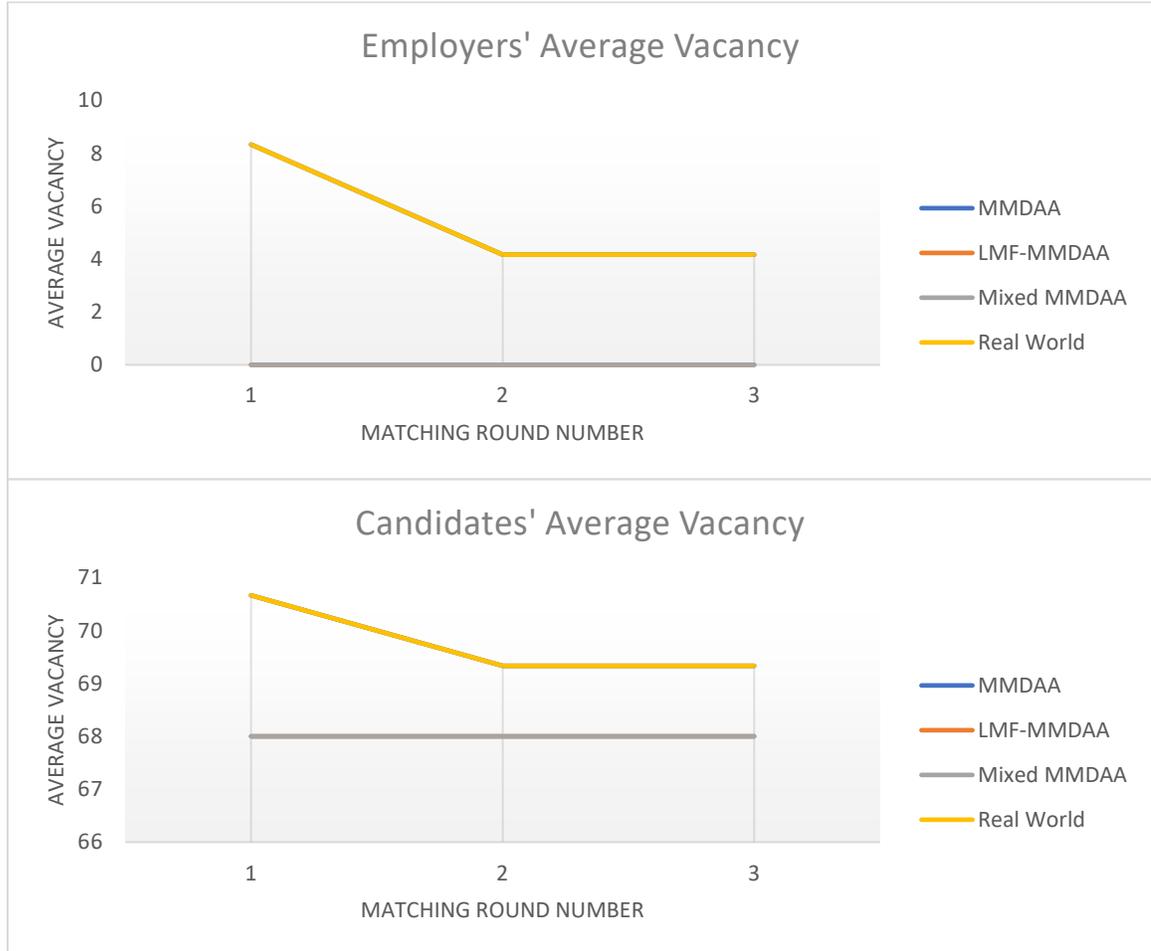

Figure 6.2: EconMatch Dataset Job Market Simulation Results

For the employers' average vacancy, the Real World simulation had a higher average vacancy compared to any of the MMDAA simulations. This is because of employers offering positions to the same candidate, while the MMDAA simulations have matches that ensure employers offers positions to unique candidates. The MMDAA simulations have a 0 average vacancy, because all 24 employers were able to find a candidate to accept their position.

The candidates' average vacancy follows a similar pattern. The Real World simulation had the highest average vacancy due to candidates not being able to find a position, since employers were offering positions to non-unique candidates. Also, since there were only 24 employers, the candidate average vacancy was high due to a surplus of candidates. In other words, at most 24 candidates would be able to find a job, which is why all 3 MMDAA simulations weren't able to reach below the 68% mark. Although this is still an improvement from the Real World simulation, there would be more interesting results across the 3 MMDAA simulations themselves, if the number of candidates and employers were more balanced.

# Chapter 7

# Runtime Complexity

## 7.1 Runtime Measurements and Results

The below figure shows the runtime of each MMDAA across the different datasets. For each MMDAA, the maximum matching rounds was set to 10, as that's where the more important matches are at. Capping the maximum matching rounds allows us to compare each of the MMDAA's fairly. LMF-MMDAA gains more preferences through non-negative matrix factorization, which means there will be many more matching rounds. This cap is necessary because we want to compare the runtimes of each algorithm based on how long it takes to achieve the same results (in this case, the results are 10 matches).

| MMDAA Runtimes | **Normal MMDAA** | **LMF-MMDAA** | **Mixed MMDAA** |
|---|---|---|---|
| **10x100 Dataset** | 13 ms | 56 ms | 91 ms |
| **50x100 Dataset** | 85 ms | 996 ms | 442 ms |
| **100x100 Dataset** | 439 ms | 10.77 sec | 118 ms |
| **110x100 Dataset** | 573 ms | 19.87 sec | 480 ms |
| **150x100 Dataset** | 1.43 sec | 30.45 sec | 2.96 sec |
| **EconMatch Dataset** | 56 ms | 1.49 sec | 88 ms |

Figure 7.1: MMDAA Runtimes

The worst-case runtime complexity for the Normal MMDAA is $O(kn^2)$, where $k$ is the number of rounds and $n$ is *max(# of candidates, # of employers)*. Since the number of rounds is constant, the worst-case runtime complexity is $O(n^2)$. For 10 matching rounds, we can see that the Normal MMDAA has a much smaller runtime compared to LMF-MMDAA across all of the datasets. The worst-case runtime complexity for LMF-MMDAA is also $O(n^2)$. LMF-MMDAA is much slower compared to the Normal MMDAA is because of the additional preferences LMF-MMDAA has to iterate through, and also break/engage. In order for these one-to-one matches to be stable, the candidates and employers must match together such that there are no two people who would be better matched with each other than their current partners. As explained in the technical section, the DAA ensures stability by initially "engaging" candidates and employers together, then iterate through everything again to ensure that the matches are stable. If they aren't stable, this "engagement" is broken and a new, more stable, one is formed. This process of engagement and breaking engagements occurs at a much higher rate in the LMF-MMDAA, as there are many more preferences to go through.

Essentially, LMF-MMDAA's runtime complexity is almost always going to be the worst case, $O(n^2)$. For Normal MMDAA, the runtime complexity is usually around $O(n*m)$, where $n$ is *max(# of candidates, # of employers)*, and $m$ is the maximum number of preferences found across all candidates or employers. In the LMF-MMDAA, $m$ is basically equal to $n$, because non-negative matrix factorization will 1) generate a preference for all candidates for each employer, and 2) generate a preference for all employers for each candidate. In other words, each candidate and employer will have a preference on each other, making *m* equal to *n*. Because of this, the runtime complexity for LMF-MMDAA will usually be O($n*m$) -> O($n*n$) -> O($n^2$).

We also see that the runtime complexity is quite low for Mixed MMDAA. Unlike Normal MMDAA and LMF-MMDAA, the Mixed MMDAA's runtime doesn't really depend on the number of preferences and process of engagement and breaking engagements, it depends on 3 parameters. The runtime complexity as mentioned in the technical section is O($n*m*k$), where *n* is the *max(# of candidates, # of employers)*, *m* is the number of matches from the Normal MMDAA, and *k* is the number of matches from the LMF-MMDAA. Both *m* and *k* will be 10 in this case, since we capped the number of matching rounds to 10. This results in the runtime complexity of the Mixed MMDAA to be O($n$).

Note that there are 2 preprocessing stages that we did not include in the runtime results of the above figure. 1) The runtime for training the LMF model, (non-negative matrix factorization). 2) In other to use the Mixed MMDAA, we must have first ran the Normal MMDAA and LMF-MMDAA to acquire the necessary matches. The runtime for the Mixed MMDAA in the above figure is solely for the Mixed MMDAA itself. These runtimes were tracked on a machine that has an Intel Core i7, 7$^{th}$ Gen. processor, with 16 GB of ram.

# Chapter 8

## Conclusion

We took a novel approach to job recommender systems by adapting the DAA into a recommender system that finds multiple stable matches between candidates and employers as recommendations for employment. Machine learning techniques, such as LMF, were used on the bipartisan preferences to enhance the resulting matches this DAA adaptation. After analyzing each of the three MMDAA's in terms of their displacement, withholdings, and vacancy, we found that the Mixed MMDAA has the best performance. Across all the experimental datasets, the Mixed MMDAA achieved the lowest average employers' and candidates' displacement compared to the Normal MMDAA and LMF-MMDAA. While LMF-MMDAA significantly lowers the amount withholdings for both employers and candidates compared to the Normal MMDAA, it still yields a high average withholdings because employers and candidates aren't getting removed out of the MMDAA due to pending preferences. The Mixed MMDAA resolves this issue since it simply fills in withholdings in the Normal MMDAA matches by using the LMF-MMDAA matches as substitutes, yielding the lowest average withholdings across all datasets. In terms of vacancy in a job market simulation, the Normal MMDAA already yields a great average vacancy when compared to a typical decentralized job market in which employers offer positions based off of their own classes and preferences. Our findings show that the Mixed MMDAA offers stable matches that maintains or even improves the average vacancy and accuracy of the Normal MMDAA. Using a dataset (EconMatch) composed of real data, rather than generated data, we were able to draw the same conclusions as described for the experimental datasets. We deployed our novel recommender system in a real application with real users and showed that the performance is more optimal in terms of job vacancy and jobless candidates than a standard decentralized job market simulation, achieving our goal of maximizing job market stability.